\documentclass[aps,prd,twocolumn,superscriptaddress,showpacs,letterpaper,floatfix]{revtex4}

\usepackage{graphicx}
\usepackage{amsmath, amsthm, amssymb}
\usepackage{booktabs}
\usepackage{color}

\begin{document}

\title{Study of orbitally excited $B$ mesons and evidence for a new $B\pi$ resonance}

\affiliation{Institute of Physics, Academia Sinica, Taipei, Taiwan 11529, Republic of China}
\affiliation{Argonne National Laboratory, Argonne, Illinois 60439, USA}
\affiliation{University of Athens, 157 71 Athens, Greece}
\affiliation{Institut de Fisica d'Altes Energies, ICREA, Universitat Autonoma de Barcelona, E-08193, Bellaterra (Barcelona), Spain}
\affiliation{Baylor University, Waco, Texas 76798, USA}
\affiliation{Istituto Nazionale di Fisica Nucleare Bologna, \ensuremath{^{jj}}University of Bologna, I-40127 Bologna, Italy}
\affiliation{University of California, Davis, Davis, California 95616, USA}
\affiliation{University of California, Los Angeles, Los Angeles, California 90024, USA}
\affiliation{Instituto de Fisica de Cantabria, CSIC-University of Cantabria, 39005 Santander, Spain}
\affiliation{Carnegie Mellon University, Pittsburgh, Pennsylvania 15213, USA}
\affiliation{Enrico Fermi Institute, University of Chicago, Chicago, Illinois 60637, USA}
\affiliation{Comenius University, 842 48 Bratislava, Slovakia; Institute of Experimental Physics, 040 01 Kosice, Slovakia}
\affiliation{Joint Institute for Nuclear Research, RU-141980 Dubna, Russia}
\affiliation{Duke University, Durham, North Carolina 27708, USA}
\affiliation{Fermi National Accelerator Laboratory, Batavia, Illinois 60510, USA}
\affiliation{University of Florida, Gainesville, Florida 32611, USA}
\affiliation{Laboratori Nazionali di Frascati, Istituto Nazionale di Fisica Nucleare, I-00044 Frascati, Italy}
\affiliation{University of Geneva, CH-1211 Geneva 4, Switzerland}
\affiliation{Glasgow University, Glasgow G12 8QQ, United Kingdom}
\affiliation{Harvard University, Cambridge, Massachusetts 02138, USA}
\affiliation{Division of High Energy Physics, Department of Physics, University of Helsinki, FIN-00014, Helsinki, Finland; Helsinki Institute of Physics, FIN-00014, Helsinki, Finland}
\affiliation{University of Illinois, Urbana, Illinois 61801, USA}
\affiliation{The Johns Hopkins University, Baltimore, Maryland 21218, USA}
\affiliation{Institut f\"{u}r Experimentelle Kernphysik, Karlsruhe Institute of Technology, D-76131 Karlsruhe, Germany}
\affiliation{Center for High Energy Physics: Kyungpook National University, Daegu 702-701, Korea; Seoul National University, Seoul 151-742, Korea; Sungkyunkwan University, Suwon 440-746, Korea; Korea Institute of Science and Technology Information, Daejeon 305-806, Korea; Chonnam National University, Gwangju 500-757, Korea; Chonbuk National University, Jeonju 561-756, Korea; Ewha Womans University, Seoul, 120-750, Korea}
\affiliation{Ernest Orlando Lawrence Berkeley National Laboratory, Berkeley, California 94720, USA}
\affiliation{University of Liverpool, Liverpool L69 7ZE, United Kingdom}
\affiliation{University College London, London WC1E 6BT, United Kingdom}
\affiliation{Centro de Investigaciones Energeticas Medioambientales y Tecnologicas, E-28040 Madrid, Spain}
\affiliation{Massachusetts Institute of Technology, Cambridge, Massachusetts 02139, USA}
\affiliation{University of Michigan, Ann Arbor, Michigan 48109, USA}
\affiliation{Michigan State University, East Lansing, Michigan 48824, USA}
\affiliation{Institution for Theoretical and Experimental Physics, ITEP, Moscow 117259, Russia}
\affiliation{University of New Mexico, Albuquerque, New Mexico 87131, USA}
\affiliation{The Ohio State University, Columbus, Ohio 43210, USA}
\affiliation{Okayama University, Okayama 700-8530, Japan}
\affiliation{Osaka City University, Osaka 558-8585, Japan}
\affiliation{University of Oxford, Oxford OX1 3RH, United Kingdom}
\affiliation{Istituto Nazionale di Fisica Nucleare, Sezione di Padova, \ensuremath{^{kk}}University of Padova, I-35131 Padova, Italy}
\affiliation{University of Pennsylvania, Philadelphia, Pennsylvania 19104, USA}
\affiliation{Istituto Nazionale di Fisica Nucleare Pisa, \ensuremath{^{ll}}University of Pisa, \ensuremath{^{mm}}University of Siena, \ensuremath{^{nn}}Scuola Normale Superiore, I-56127 Pisa, Italy, \ensuremath{^{oo}}INFN Pavia, I-27100 Pavia, Italy, \ensuremath{^{pp}}University of Pavia, I-27100 Pavia, Italy}
\affiliation{University of Pittsburgh, Pittsburgh, Pennsylvania 15260, USA}
\affiliation{Purdue University, West Lafayette, Indiana 47907, USA}
\affiliation{University of Rochester, Rochester, New York 14627, USA}
\affiliation{The Rockefeller University, New York, New York 10065, USA}
\affiliation{Istituto Nazionale di Fisica Nucleare, Sezione di Roma 1, \ensuremath{^{qq}}Sapienza Universit\`{a} di Roma, I-00185 Roma, Italy}
\affiliation{Mitchell Institute for Fundamental Physics and Astronomy, Texas A\&M University, College Station, Texas 77843, USA}
\affiliation{Istituto Nazionale di Fisica Nucleare Trieste, \ensuremath{^{rr}}Gruppo Collegato di Udine, \ensuremath{^{ss}}University of Udine, I-33100 Udine, Italy, \ensuremath{^{tt}}University of Trieste, I-34127 Trieste, Italy}
\affiliation{University of Tsukuba, Tsukuba, Ibaraki 305, Japan}
\affiliation{Tufts University, Medford, Massachusetts 02155, USA}
\affiliation{University of Virginia, Charlottesville, Virginia 22906, USA}
\affiliation{Waseda University, Tokyo 169, Japan}
\affiliation{Wayne State University, Detroit, Michigan 48201, USA}
\affiliation{University of Wisconsin, Madison, Wisconsin 53706, USA}
\affiliation{Yale University, New Haven, Connecticut 06520, USA}

\author{T.~Aaltonen}
\affiliation{Division of High Energy Physics, Department of Physics, University of Helsinki, FIN-00014, Helsinki, Finland; Helsinki Institute of Physics, FIN-00014, Helsinki, Finland}
\author{S.~Amerio\ensuremath{^{kk}}}
\affiliation{Istituto Nazionale di Fisica Nucleare, Sezione di Padova, \ensuremath{^{kk}}University of Padova, I-35131 Padova, Italy}
\author{D.~Amidei}
\affiliation{University of Michigan, Ann Arbor, Michigan 48109, USA}
\author{A.~Anastassov\ensuremath{^{w}}}
\affiliation{Fermi National Accelerator Laboratory, Batavia, Illinois 60510, USA}
\author{A.~Annovi}
\affiliation{Laboratori Nazionali di Frascati, Istituto Nazionale di Fisica Nucleare, I-00044 Frascati, Italy}
\author{J.~Antos}
\affiliation{Comenius University, 842 48 Bratislava, Slovakia; Institute of Experimental Physics, 040 01 Kosice, Slovakia}
\author{G.~Apollinari}
\affiliation{Fermi National Accelerator Laboratory, Batavia, Illinois 60510, USA}
\author{J.A.~Appel}
\affiliation{Fermi National Accelerator Laboratory, Batavia, Illinois 60510, USA}
\author{T.~Arisawa}
\affiliation{Waseda University, Tokyo 169, Japan}
\author{A.~Artikov}
\affiliation{Joint Institute for Nuclear Research, RU-141980 Dubna, Russia}
% ...

\collaboration{CDF Collaboration}
\altaffiliation[With visitors from]{
\ensuremath{^{a}}University of British Columbia, Vancouver, BC V6T 1Z1, Canada,
\ensuremath{^{b}}Istituto Nazionale di Fisica Nucleare, Sezione di Cagliari, 09042 Monserrato (Cagliari), Italy,
\ensuremath{^{c}}University of California Irvine, Irvine, CA 92697, USA,
\ensuremath{^{d}}Institute of Physics, Academy of Sciences of the Czech Republic, 182~21, Czech Republic,
\ensuremath{^{e}}CERN, CH-1211 Geneva, Switzerland,
\ensuremath{^{f}}Cornell University, Ithaca, NY 14853, USA,
\ensuremath{^{g}}University of Cyprus, Nicosia CY-1678, Cyprus,
\ensuremath{^{h}}Office of Science, U.S. Department of Energy, Washington, DC 20585, USA,
\ensuremath{^{i}}University College Dublin, Dublin 4, Ireland,
\ensuremath{^{j}}ETH, 8092 Z\"{u}rich, Switzerland,
\ensuremath{^{k}}University of Fukui, Fukui City, Fukui Prefecture, Japan 910-0017,
\ensuremath{^{l}}Universidad Iberoamericana, Lomas de Santa Fe, M\'{e}xico, C.P. 01219, Distrito Federal,
\ensuremath{^{m}}University of Iowa, Iowa City, IA 52242, USA,
\ensuremath{^{n}}Kinki University, Higashi-Osaka City, Japan 577-8502,
\ensuremath{^{o}}Kansas State University, Manhattan, KS 66506, USA,
\ensuremath{^{p}}Brookhaven National Laboratory, Upton, NY 11973, USA,
\ensuremath{^{q}}University of Manchester, Manchester M13 9PL, United Kingdom,
\ensuremath{^{r}}Queen Mary, University of London, London, E1 4NS, United Kingdom,
\ensuremath{^{s}}University of Melbourne, Victoria 3010, Australia,
\ensuremath{^{t}}Muons, Inc., Batavia, IL 60510, USA,
\ensuremath{^{u}}Nagasaki Institute of Applied Science, Nagasaki 851-0193, Japan,
\ensuremath{^{v}}National Research Nuclear University, Moscow 115409, Russia,
\ensuremath{^{w}}Northwestern University, Evanston, IL 60208, USA,
\ensuremath{^{x}}University of Notre Dame, Notre Dame, IN 46556, USA,
\ensuremath{^{y}}Universidad de Oviedo, E-33007 Oviedo, Spain,
\ensuremath{^{z}}CNRS-IN2P3, Paris, F-75205 France,
\ensuremath{^{aa}}Universidad Tecnica Federico Santa Maria, 110v Valparaiso, Chile,
\ensuremath{^{bb}}The University of Jordan, Amman 11942, Jordan,
\ensuremath{^{cc}}Universite catholique de Louvain, 1348 Louvain-La-Neuve, Belgium,
\ensuremath{^{dd}}University of Z\"{u}rich, 8006 Z\"{u}rich, Switzerland,
\ensuremath{^{ee}}Massachusetts General Hospital, Boston, MA 02114 USA,
\ensuremath{^{ff}}Harvard Medical School, Boston, MA 02114 USA,
\ensuremath{^{gg}}Hampton University, Hampton, VA 23668, USA,
\ensuremath{^{hh}}Los Alamos National Laboratory, Los Alamos, NM 87544, USA,
\ensuremath{^{ii}}Universit\`{a} degli Studi di Napoli Federico I, I-80138 Napoli, Italy
}
\noaffiliation

\begin{abstract}
Using the full CDF Run II data sample, we report evidence for a new resonance, which we refer to as $B(5970)$, found simultaneously
in the $B^0\pi^+$ and $B^+\pi^-$ mass distributions with a significance of 4.4 standard deviations.
We further report the first study of resonances consistent with
orbitally excited $B^{+}$ mesons and
an updated measurement of the properties of
orbitally excited $B^0$ and $B_s^0$ mesons.
We measure the masses and widths of all states,
as well as 
the relative production rates of the $B_1$, $B_2^*$, and $B(5970)$ states and 
the branching fraction of the 
$B_{s2}^{*0}$ state to either $B^{*+} K^-$ and $B^{+} K^-$.
Furthermore, we measure the production rates of the orbitally excited $B^{0,+}$ states relative to the $B^{0,+}$ ground state.
The masses of the new $B(5970)$ resonances are
$5978\pm5(\textrm{stat})\pm12(\textrm{syst}) \textrm{ MeV/c}^{2}$ for the neutral state and $5961\pm5(\textrm{stat})\pm12(\textrm{syst}) \textrm{ MeV/c}^{2}$ for the charged state, assuming that the resonance decays into $B\pi$ final states.
The properties of the orbitally excited and the new $B(5970)^{0,+}$ states are 
compatible with isospin symmetry.

\end{abstract}

\maketitle

\setcounter{secnumdepth}{2}
\renewcommand{\thetable}{\Roman{table}}

\newcommand{\BDP}{$B^{+} \rightarrow \overline{D}^{0} \pi^{+}$}
\newcommand{\BDd}{$B^{+} \rightarrow \overline{D}^{0} (\pi^{+}\pi^{-})\pi^{+}$}
\newcommand{\BJ}{$B^{+} \rightarrow J/\psi K^{+}$}
\newcommand{\BzDP}{$B^{0} \rightarrow D^{-} \pi^{+}$}
\newcommand{\BzDd}{$B^{0} \rightarrow D^{-} (\pi^{+}\pi^{-})\pi^{+}$}
\newcommand{\BzJS}{$B^{0} \rightarrow J/\psi K^{*0}$}
\newcommand{\BzJs}{$B^{0} \rightarrow J/\psi K_{S}$}

\newcommand{\BDPB}{$\boldsymbol{B^{+} \rightarrow \overline{D}^{0} \pi^{+}}$}
\newcommand{\BDdB}{$\boldsymbol{B^{+} \rightarrow \overline{D}^{0} (\pi^{+}\pi^{-})\pi^{+}}$}
\newcommand{\BJB}{$\boldsymbol{B^{+} \rightarrow J/\psi K^{+}}$}
\newcommand{\BzDPB}{$\boldsymbol{B^{0} \rightarrow D^{-} \pi^{+}}$}
\newcommand{\BzDdB}{$\boldsymbol{B^{0} \rightarrow D^{-} (\pi^{+}\pi^{-})\pi^{+}}$}
\newcommand{\BzJSB}{$\boldsymbol{B^{0} \rightarrow J/\psi K^{*0}}$}
\newcommand{\BzJsB}{$\boldsymbol{B^{0} \rightarrow J/\psi K_{S}}$}

\newcommand{\sPlot}{$_s\mathcal{P}lot\, $}

\newcommand{\ws}{}
\newcommand{\bs}{$<$}

\
\\
\section{Introduction}

The detailed study of hydrogen atom emission spectra was essential for the understanding of quantum electrodynamics.
This is partially due to the simple composition of the hydrogen atom,
consisting of just two particles, and partially due to the large mass difference
between the proton and the electron, which mostly decouples the proton spin from the electron spin.
As a consequence, the fine- and hyperfine structures of hydrogen atoms are characterized by significantly different energy scales.
Similarly, the detailed study of mesons composed of a heavy and a light valence quark supports the understanding of quantum chromodynamics
and the limitations of its low-energy approximations, such as the heavy quark effective theory (HQET) \cite{HQET}.
The spectroscopy of $B_{(s)}$ mesons, which contain a $\overline{b}$ quark and a
$u$ or $d$ (or $s$) quark, provides an important testing ground for HQET. 

The ground state $B_{(s)}$ mesons and the spin-1 $B^*_{(s)}$ mesons have been thoroughly studied \cite{PDBook}.
This paper studies the states with
orbital angular momentum $L = 1$ and a higher excited state.
For each type of $B$ meson, four distinct states with $L = 1$ are possible,
each with different
couplings between the spin of the quarks and the orbital angular momentum.
Assuming the bottom quark to be heavy, HQET predicts that the dynamics is dominated by the coupling between the orbital angular momentum
and the spin of the light quark that combine to a total light-quark angular momentum $j = \frac{1}{2}$ or 
$j = \frac{3}{2}$, which corresponds to the fine structure in the hydrogen atom.
Additional contributions arise due to the spin of the $\overline{b}$ quark. This results in two doublets of states,
corresponding to fine- and hyperfine-splitting, 
that are collectively referred to as $B^{**}_{(s)}$ mesons.
The states with $j = \frac{1}{2}$ are named
$B_0^*$ ($J = 0$) and 
$B_1$ ($J = 1$) mesons; the states with $j = \frac{3}{2}$ are named
$B_1$ ($J = 1$) and 
$B_2^*$ ($J = 2$) mesons, where $J$ is the total angular momentum.

In HQET, different results originate from various
approximations adopted in the calculation of the light-quark degrees of freedom.
Such calculations can neglect or include relativistic effects
as well as the dynamical spin dependence of the potential between the quarks.
While most of the recent predictions are based on HQET~\cite{hqet1,hqet2,hqet3,hqet4,hqet6}, other approaches exist, including predictions using 
lattice-gauge calculations \cite{lattice1,lattice2}, potential models \cite{potmod1,potmod2},
heavy quark symmetry (HQS) \cite{hqs}, chiral theory \cite{chiral2}, and QCD strings \cite{qcdstring}, allowing the masses, widths, and relative branching ratios to be calculated.
Predictions of $B_{(s)}^{**}$ properties are shown in Tables~\ref{tab:PredictedMasses} and \ref{tab:PredictedWidths}.

\begin{table}[h]
     \caption{Predicted $B_{(s)}^{**}$ masses. All values are in MeV$/c^2$.}
     \label{tab:PredictedMasses}
     \centering 
    \begin{tabular}{cccccc}
      \hline \hline
      Calculation & Ref. & $B_1^{0,+}$ & $B_2^{*0,+}$ & $B_{s1}^0$ & $B_{s2}^{*0}$ \\
      \hline
      HQET & \cite{hqet1} & 5700 & 5715 &  &  \\
      HQET & \cite{hqet2} & $5780 \pm 40$ & $5794 \pm 40$ & $5886 \pm 40$ & $5899 \pm 49$ \\
      HQET & \cite{hqet3} & 5623 & 5637 & 5718 & 5732 \\
      HQET & \cite{hqet4} & 5720 & 5737 & 5831 & 5847 \\
      HQET & \cite{hqet6} & 5719 & 5733 & 5831 & 5844 \\
      Lattice & \cite{lattice1} & $5732 \pm 33$ & $5772 \pm 29$ & $5815 \pm 22$ & $5845 \pm 21$ \\
      Lattice & \cite{lattice2} &  &  & $5892 \pm 52$ & $5904 \pm 52$ \\
      Potential & \cite{potmod1} & 5699 & 5704 & 5805 & 5815 \\
      Potential & \cite{potmod2} & 5780 & 5800 & 5860 & 5880 \\
      HQS & \cite{hqs} & 5755 & 5767 & 5834 & 5846 \\
      Chiral theo. & \cite{chiral2} & $5774 \pm 2$ & $5790 \pm 2$ & $5877 \pm 3$ & $5893 \pm 3$ \\
      QCD string & \cite{qcdstring} & 5716 & 5724 &  &  \\
      \hline \hline
     \end{tabular}
\end{table}

\begin{table}[h]
     \caption{Predicted $B_{(s)}^{**}$ widths. All values are in MeV$/c^2$.}
     \label{tab:PredictedWidths}
     \centering 
    \begin{tabular}{ccccc}
      \hline \hline
      Ref. & $B_1^{0,+}$ & $B_2^{*0,+}$ & $B_{s1}^0$ & $B_{s2}^{*0}$ \\
      \hline
      \cite{hqet2} &  & $16 \pm 5$ & $2.8 \pm 1.2$ & $7 \pm 3$ \\
      \cite{hqet3} & 20 & 29 &  &  \\
      \cite{potmod2} &  & 27 &  & 1.9 \\
      \cite{hqs} & 31 -- 55 & 38 -- 63 & 1 -- 3 & 3 -- 7 \\
      \cite{chiral2} & $43 \pm 10$ & $57.3 \pm 13.5$ & $3.5 \pm 1.0$ & $11.3 \pm 2.6$ \\
      \hline \hline
     \end{tabular}
\end{table}

The $B^{**0,+}$ states with $j=\frac{1}{2}$ can decay to $B^{(*)}\pi$ final states via
an $S$-wave transition and therefore are expected to 
be too broad to be distinguishable from background at current experiments, while the $j=\frac{3}{2}$ states decay via a $D$-wave.
Decays via $P$-wave are incompatible with parity conservation, as $B^{**}_{(s)}$ states have positive parity.

\begin{figure}[h]
	\centering
	\includegraphics[scale=0.33]{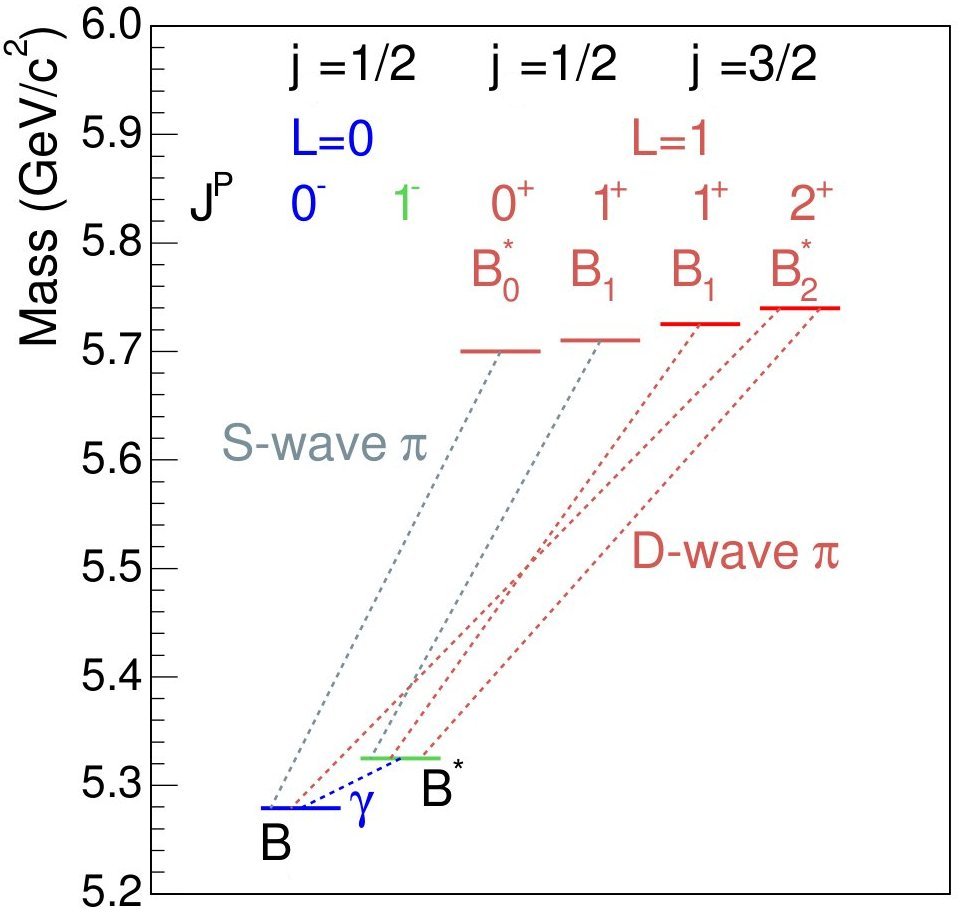}
	\caption{Spectrum and allowed decays for the lowest orbitally excited states $B^{**0,+}$. For $B^{**0}_s$ mesons the pion is replaced by a kaon and the states have higher masses.}
	\label{fig:spec}
\end{figure}

As the $B_{2}^{*}$ can decay either to $B \pi$ or $B^{*} \pi$ final states, and the 
low-energy photon from the $B^* \rightarrow B \gamma$ 
decay is typically not reconstructed, the decays of this state yield
two structures in the $B \pi$ invariant mass spectrum.
The orbital excitations of $B^0_s$ mesons are expected to have the same phenomenology as those of $B^{0,+}$ mesons. 
They decay to $B^0 \bar{K}^0$ and $B^+ K^-$ final states, but not to $B^0_s \pi^0$, due to isospin conservation in the strong-interaction decay.
Throughout this paper, charge conjugate states are implied.
The spectrum and possible decays of $B^{**0,+}$ mesons are illustrated in Fig.~\ref{fig:spec}.

Orbitally excited $B$ mesons were first observed in electron-positron collisions at LEP in 1995~\cite{lep1,lep2,lep3,lep4}.
Tevatron experiments in proton-antiproton collisions observed three structures in the $B^0 \pi^+$ invariant-mass 
distribution that were associated with the $j = \frac{3}{2}$ $B^{**0}$ meson states in the HQET approximation.
A $2.8 \sigma$ discrepancy is observed between measurements of the mass difference of the $B_2^{*0}$ and $B_1^0$ states by the D0~\cite{D0mea} and CDF Collaborations~\cite{CDFmea} using $1.3$~fb$^{-1}$ and $1.7$~fb$^{-1}$ of data, respectively.
While CDF measured $\Delta m(B^{**0}) = m_{B_{2}^{*0}}-m_{B_{1}^{0}} = 14.9^{+2.2}_{-2.5} \textrm{(stat)} ^{+1.2}_{-1.4} \textrm{(syst)}$~MeV$/c^{2}$, D0 found $\Delta m(B^{**0}) = 26.3\pm3.1 \textrm{(stat)} \pm0.9 \textrm{(syst)}$~MeV$/c^{2}$.

The $B_{s1}^0$ state was discovered by CDF \cite{CDFsmea} using $1$~fb$^{-1}$ of data.
The decay of the $B_{s2}^{*0}$ state to a $B^+ K^-$ final state was first observed by CDF \cite{CDFsmea} and D0 \cite{D0smea}, while the $B^{*+} K^-$ decay was only recently observed by LHCb \cite{LHCbmea}.
Charged $B^{**+}$ states have not been observed so far.
Preliminary measurements of $B^{**0,+}$  properties were reported by LHCb \cite{LHCbconf}.

This paper reports measurements of masses, natural widths, and relative production rates of orbitally excited  $B^{**0}$, $B^{**+}$, and $B^{**0}_s$ mesons.
For rate measurements
we define the product of the $B_1$ production rate relative to the $B_2^*$ rate times the branching fractions of the observed decays,
\begin{equation}
r_\mathrm{prod} = \frac{\sigma (B_{1})}{\sigma (B_{2}^{*})} \cdot \frac{\mathcal{B}(B_{1} \rightarrow B^{*} h)}
		{ \mathcal{B}(B_{2}^{*} \rightarrow B h) + \mathcal{B}(B_{2}^{*} \rightarrow B^{*} h) },
\label{eqn:rprod}
\end{equation}
where $\sigma$ is the production cross-section restricted to the relevant kinematic regime, and $h$ identifies $\pi$ for 
$B^{**0,+}$ and $K$ for $B_s^{**0}$ decays.
We also define the relative $B_{s2}^{*}$ branching fraction
\begin{equation}
r_\mathrm{dec}=\frac{\mathcal{B}(B_{s2}^{*} \rightarrow B^{*+} K^{-})}{\mathcal{B}(B_{s2}^{*} \rightarrow B^{+} K^{-})}.
\label{eqn:rdec}
\end{equation}
Ground-state $B$ mesons are reconstructed in seven different decay modes and combined with an additional pion (kaon) to form $B^{**}_{(s)}$ candidates.
Selections based on artificial neural networks are performed
to enrich the $B^{**}_{(s)}$ signal fractions in the samples.
The properties of the $B^{**}_{(s)}$ states are determined from fits to mass difference spectra.

\section{Data sample and event selection}

We use data from $p\overline{p}$ collisions at $\sqrt{s}$ = 1.96 TeV recorded by the CDF~II detector at the Fermilab Tevatron corresponding to the full Run II integrated luminosity of 9.6~fb$^{-1}$.
The key components of the CDF~II detector~\cite{CDFII} for these measurements 
are the charged-particle trajectory (tracking) subdetectors located in a uniform axial magnetic field
of 1.4 T, together with the muon detectors.
A single-sided silicon-strip detector mounted directly on the beam pipe at 1.5 cm radius and six layers of
double-sided silicon strips extending to a radius of 22 cm \cite{det18} 
provide a resolution of approximately 40~$\mu$m on the impact
parameter, defined as the distance between the interaction point and the trajectory of a charged particle,
projected into the plane transverse to the beam. This includes a 32~$\mu$m contribution from the transverse beam size~\cite{det18}.
An open-cell drift chamber, which covers a radius range of 45 to 137 cm  \cite{det16},
allows precise measurement of the momentum of charged particles with a resolution of $\sigma(p_T)/p_T^2 \approx 0.1\%$/(GeV/$c$).
Outside the tracking detectors, time-of-flight detectors, and calorimeters, muons are detected in planes of drift tubes and scintillators \cite{det17}.
Charged-particle identification information is obtained from the ionization energy deposition
in the drift chamber and the measurement of the flight time of particles~\cite{tof,CDFwick}.

A three-layer online event-selection system (trigger) is implemented
in hardware and software.
 Recording of the events used in this measurement is initiated by two types of triggers, 
a $J/\psi$ trigger~\cite{det14} and a displaced-track trigger~\cite{det20}.
The $J/\psi$ trigger is designed to record events enriched in $J/\psi \rightarrow \mu^{+} \mu^{-}$ decays and requires two tracks in the drift chamber geometrically matched to track segments in the muon detectors.
The particles must have opposite charge; a transverse momentum $p_T$ larger than $1.5$ or $2.0$~GeV/$c$, depending on subdetector and data taking period; an azimuthal opening angle below $135^\circ$; and a dimuon mass compatible with the known $J/\psi$-meson mass.
The displaced-track trigger requires two tracks 
with impact parameters typically between 0.12 to 1~mm,
a luminosity-dependent lower threshold on the scalar sum of transverse momenta of typically 
4.5 to 6.5~GeV/$c$, and an intersection point displaced at least 0.2~mm from the primary-interaction point in the transverse plane.
These criteria preferentially select events with decays of long-lived hadrons.

Tracks are reconstructed with a pion mass hypothesis accounting for multiple scattering and energy loss. 
In the first step of the analysis, we refit them also under the kaon-mass hypothesis.
Combinations of two or three tracks constrained to originate from the same space point are formed to reconstruct
$ J/\psi \rightarrow \mu^{+} \mu^{-} $, $ \overline{D}^{0} \rightarrow K^{+} \pi^{-} $,
$ D^{-} \rightarrow K^{+} \pi^{-} \pi^{-} $, $ K^{*}(892)^{0} \rightarrow K^{+} \pi^{-} $,
and $ K_{S}^0 \rightarrow \pi^{+} \pi^{-} $ decays, where the $J/\psi$ and $\overline{D}^{0}$ candidate masses are constrained to their known values~\cite{PDBook}.
Next, $B$ mesons are formed in the following seven decay modes:
$ B^{+} \rightarrow J/\psi\, K^{+} $, $ B^{+} \rightarrow \overline{D}^{0} \pi^{+} $,
$ B^{+} \rightarrow \overline{D}^{0} (\pi^{+} \pi^{-}) \pi^{+} $,
$ B^{0} \rightarrow J/\psi\, K^{*}(892)^{0} $, $ B^{0} \rightarrow J/\psi\, K_{S}^0 $,
$ B^{0} \rightarrow D^{-} \pi^{+} $,
and $ B^{0} \rightarrow D^{-} (\pi^{+} \pi^{-}) \pi^{+} $.
Finally, we reconstruct $B^{**}_{(s)}$ mesons in the $B^{**0} \rightarrow B^{(*)+} \pi^-$, $B^{**+} \rightarrow B^{(*)0} \pi^+$ and $B^{**}_s \rightarrow B^{(*)+} K^-$ channels.
Because the photon from the $B^{*} \rightarrow B \gamma$ decay is too low in energy to be detected, $B^{*}$ mesons are partially reconstructed as $B$ mesons.
This reduces the reconstructed $B^{**}_{(s)}$ mass by approximately 46~MeV$/c^2$, the mass difference between $B^*$ and $B$ mesons.
To improve the mass resolution, we use the $Q$ value, defined as $Q=m(Bh)-m(B)-m_{h}$ instead of $m(Bh)$ to determine the resonance parameters because it reduces the effect of the $B$ reconstruction resolution.

\begin{figure}[b]
	\centering
	\includegraphics[width=0.47\textwidth]{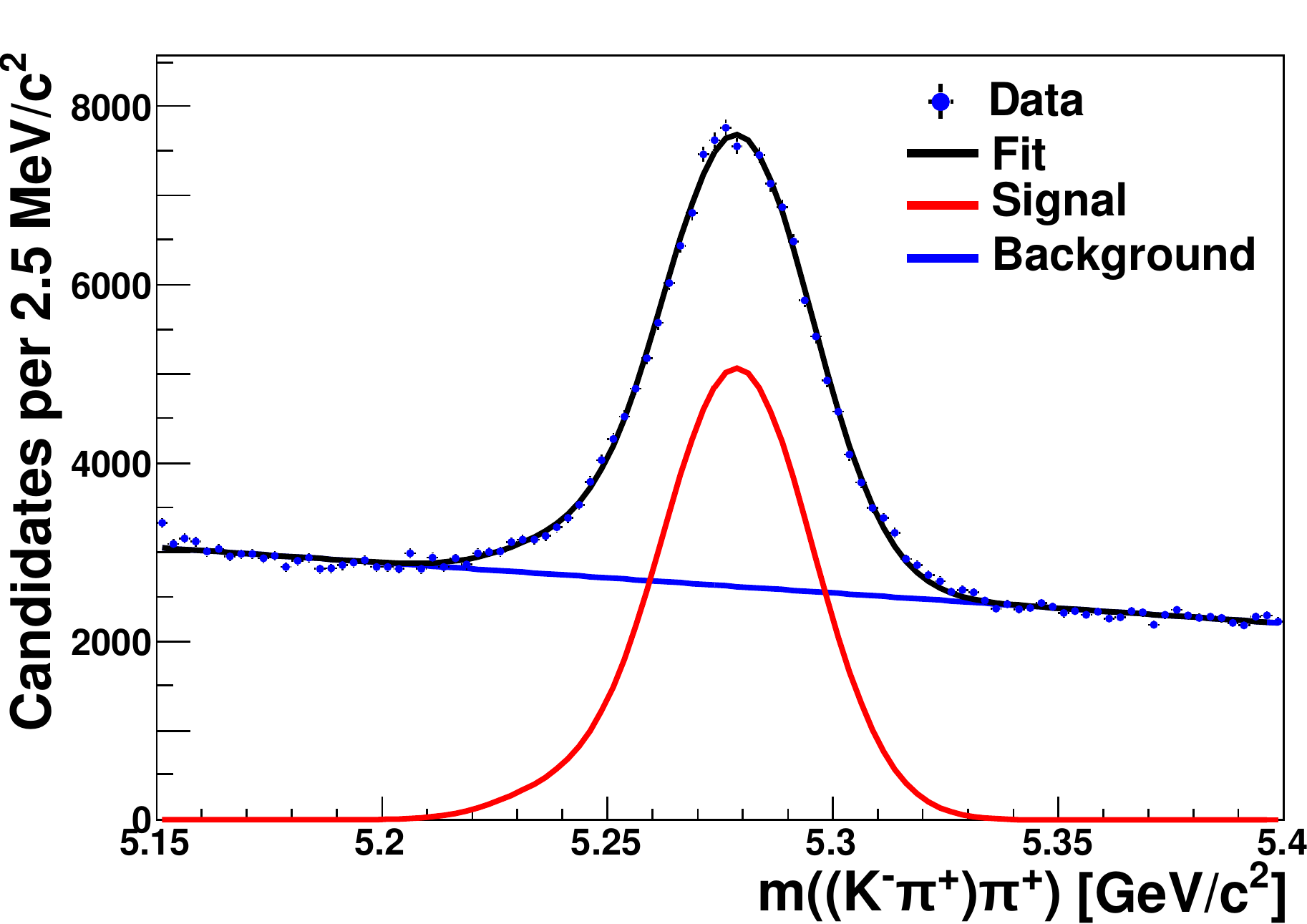}
	\caption{Invariant $K\pi\pi$-mass distribution of $B^{+} \rightarrow \overline{D}^{0}(\rightarrow  K^- \pi^{+}) \pi^{+}$ candidates after the application of loose requirements, before neural network selection, with fit result overlaid.}
	\label{fig:BMpre}
\end{figure}

Because the various $B$-meson decay channels have differing topologies, we optimize the selection separately for each channel.
First, we apply modest requirements on quantities providing significant signal-to-background separation, such as transverse momentum, transverse flight length, impact parameter, and vertex fit quality of the $B$ candidate; and transverse momenta of the final-state particles, so that $B$ meson signals become visible in the mass spectra.
An example is shown in Fig.~\ref{fig:BMpre}.
The resulting mass distributions are then fit with a linear or exponential background model and one or two Gaussians as a signal model, depending on the $B$ decay mode.
The absolute numbers of signal and background candidates, as well as the distributions as a function of $m(B)$ for signal and background, are derived from the fit.
This information is used to calculate $_s\mathcal{P}lot\, $ weights \cite{splot}.
When applied to distributions of quantities that are not correlated with $m(B)$, these weights allow the extraction of statistically-pure distributions of these quantities for signal and background separately.
Observed events and their weights are input to a multivariate classifier \cite{nb}, allowing training based on data only.
Topological, kinematic, and particle identification quantities of the $B$ mesons and their final-state particles are used as input variables.
Due to the lifetime of the $B$ mesons, the variables with the 
most discriminating power are flight length, impact parameter, and vertex-fit quality of the $B$-meson candidate.
Additional inputs are the transverse momenta and particle identification information of pions, kaons, and muons and invariant masses of intermediate decay products such as $D$ and $J/\psi$ mesons.
A moderate requirement is applied on the discriminator's output to remove candidates formed using random combination of tracks that meet the candidate's selection requirements. 
For the data set shown in Fig.~\ref{fig:BMpre}, this requirement rejects 74\% of the background while retaining 97\% of the signal.
In addition, the information from the discriminator's output is further used in the $B^{**}_{(s)}$ selection.

For the optimization of the selection of $B^{**}_{(s)}$ mesons, we rely on simulations of $B^{**}_{(s)}$ decays with the full CDF~II detector geometry.
The primary $B^{**}_{(s)}$ particle is generated using measured $b$-hadron kinematic distributions~\cite{CDFII}.
Its decay to $B^{(*)}h$ with $h=\pi, K$ and the subsequent $B$-meson decay are simulated with {\sc EvtGen}~\cite{evtgen}. 
The detector is simulated with GEANT~\cite{geant}.

Neural networks are trained to separate $B^{**}_{(s)}$ signal from background using simulations as signal and $B^{**}_{(s)}$ candidates observed in data, which contain a negligible signal fraction, as background. 
Only quantities of the $B^{**}_{(s)}$ meson and the additional pion or kaon and ground-state $B$ meson mass are used as discriminating variables. 
To avoid biasing the training to a certain mass range, simulated events are generated with the same $Q$-value distribution as the background in data.

\begin{figure}[b]
  \centering 
  \includegraphics[width=0.47\textwidth]{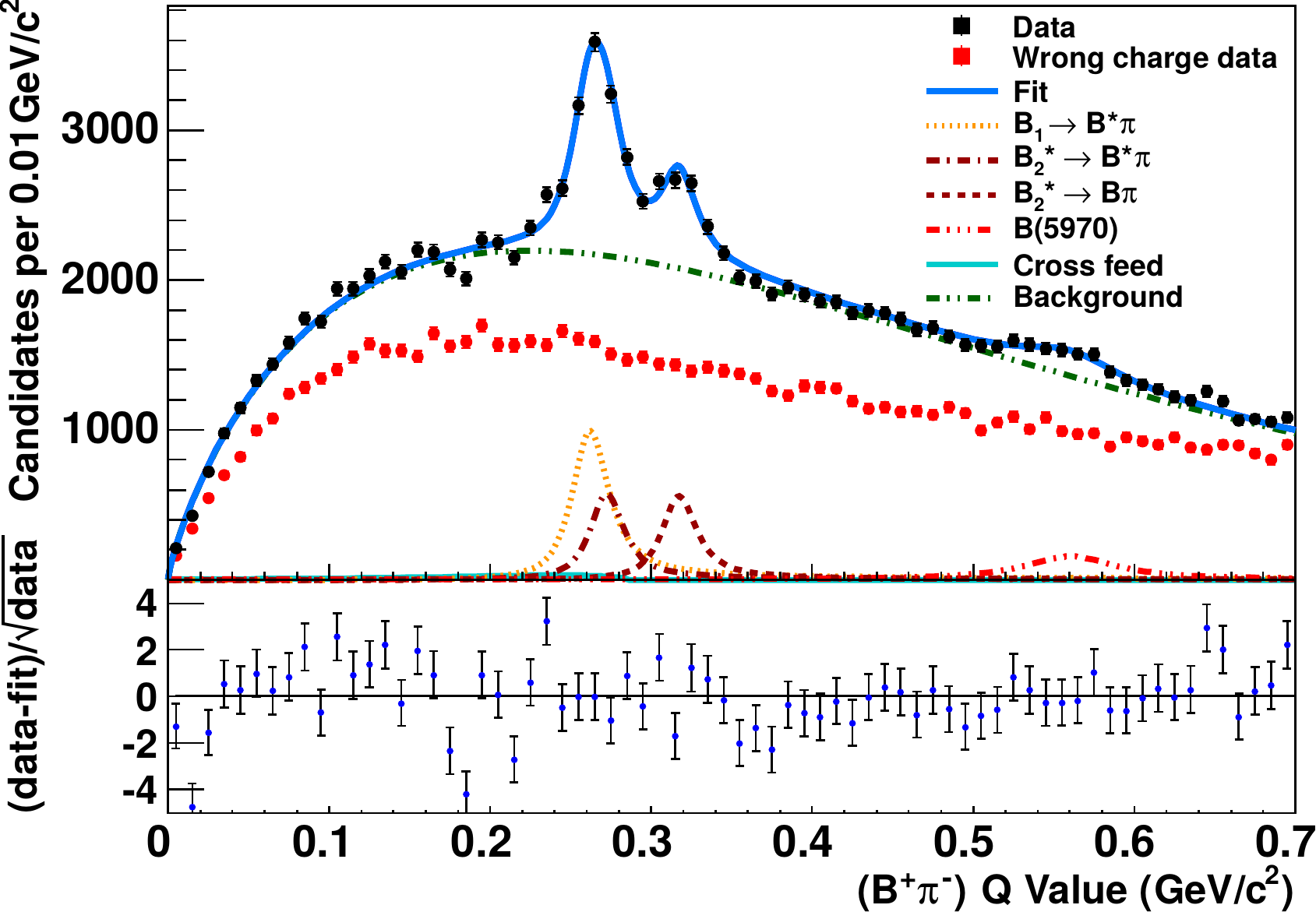} \\
  \vspace*{0.5cm} 
  \includegraphics[width=0.47\textwidth]{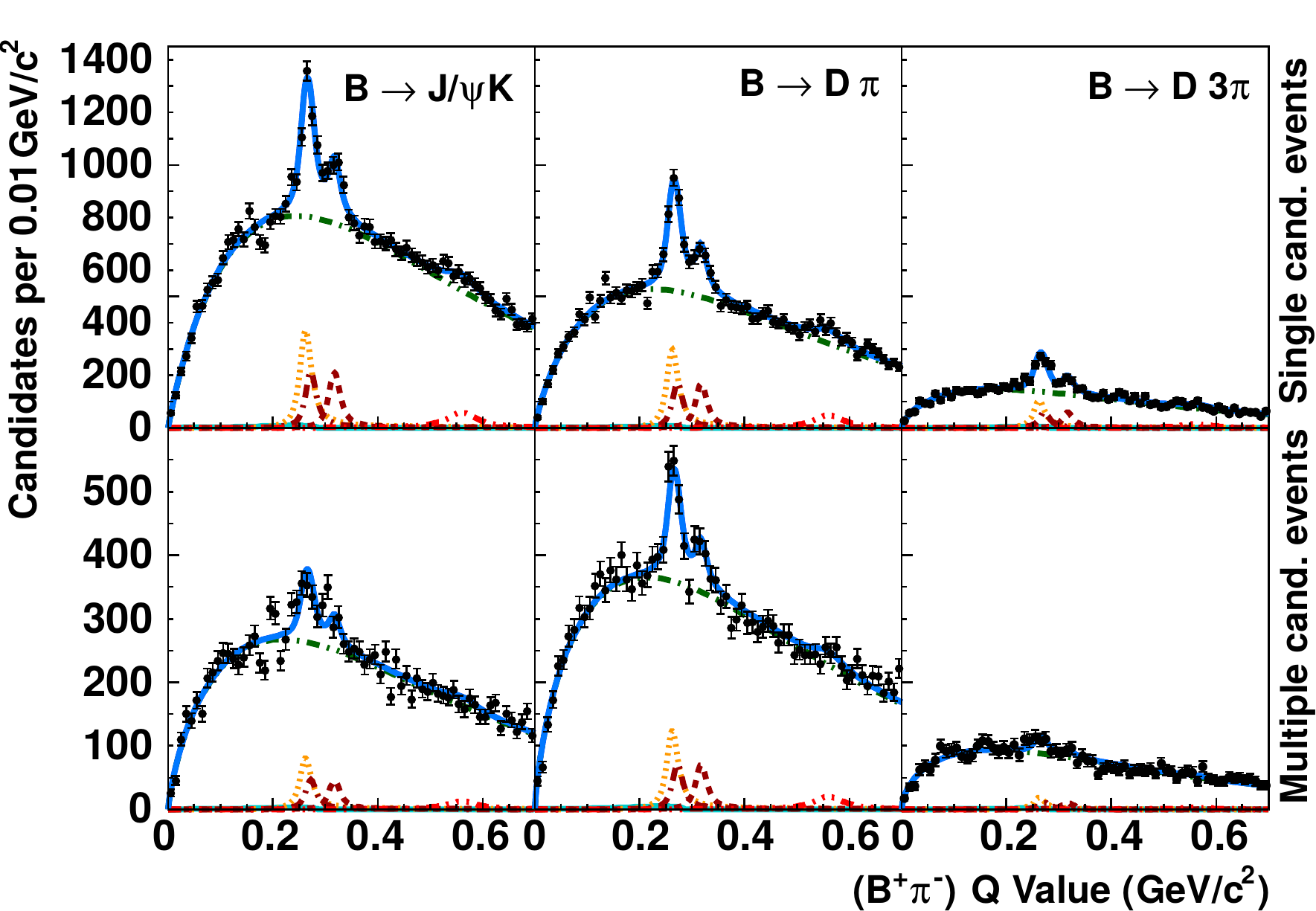}  
  \caption{Distribution of $Q$ value of $B^{**0}$ candidates (and $B^+\pi^+$ combinations in the upper plot) with fit results overlaid.
The upper panel shows the data summed over decay channels and the deviations of these from the fit function,
normalized to the poisson uncertainty of the data. The lower panels show data and fits for each decay channel individually, 
separated into events with one candidate (upper row) and with multiple candidates (lower row).}
	\label{fig:B0SSfit}
\end{figure}

\begin{figure}[h]
  \centering  
  \includegraphics[width=0.47\textwidth]{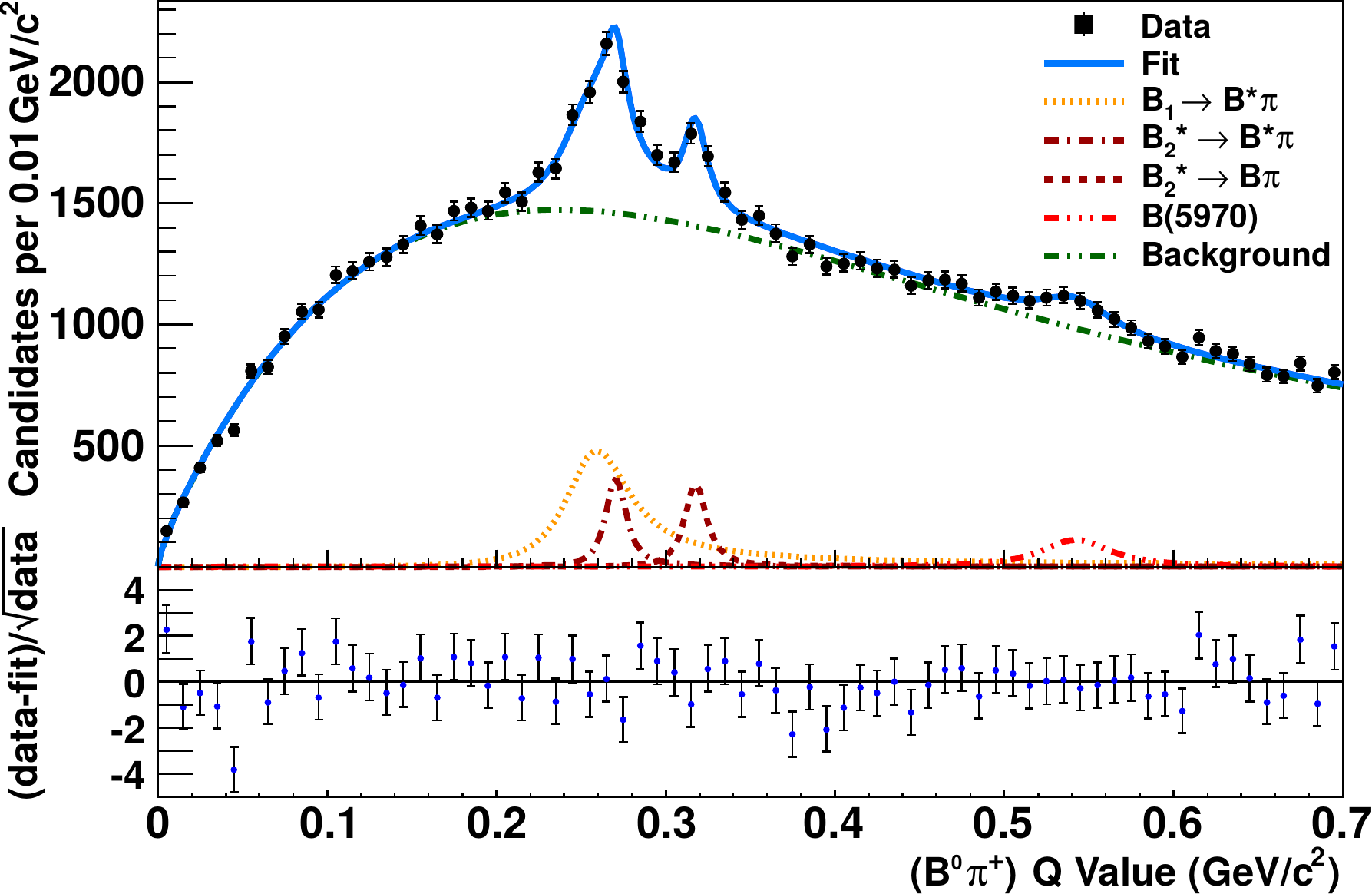} \\
  \vspace*{0.5cm} 
  \includegraphics[width=0.47\textwidth]{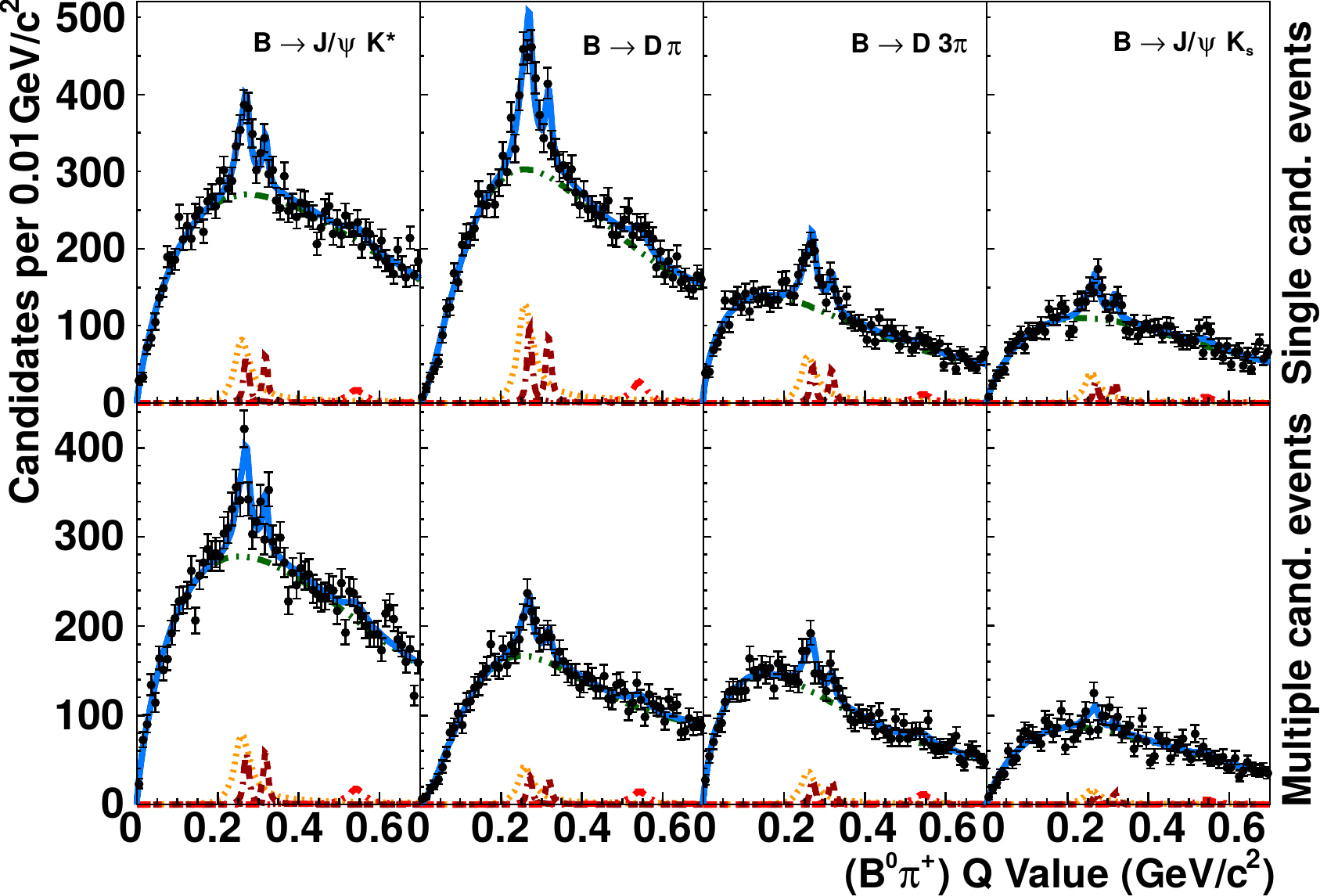}  
  \caption{Distribution of $Q$ value of $B^{**+}$ candidates with fit results overlaid.
The upper panel shows the data summed over decay channels and the deviations of these from the fit function,
normalized to the poisson uncertainty of the data. The lower panels show data and fits for each decay channel individually, 
separated into events with one candidate (upper row) and with multiple candidates (lower row).}
	\label{fig:BpSSfit}
\end{figure}

\begin{figure}[h]
  \centering  
  \includegraphics[width=0.47\textwidth]{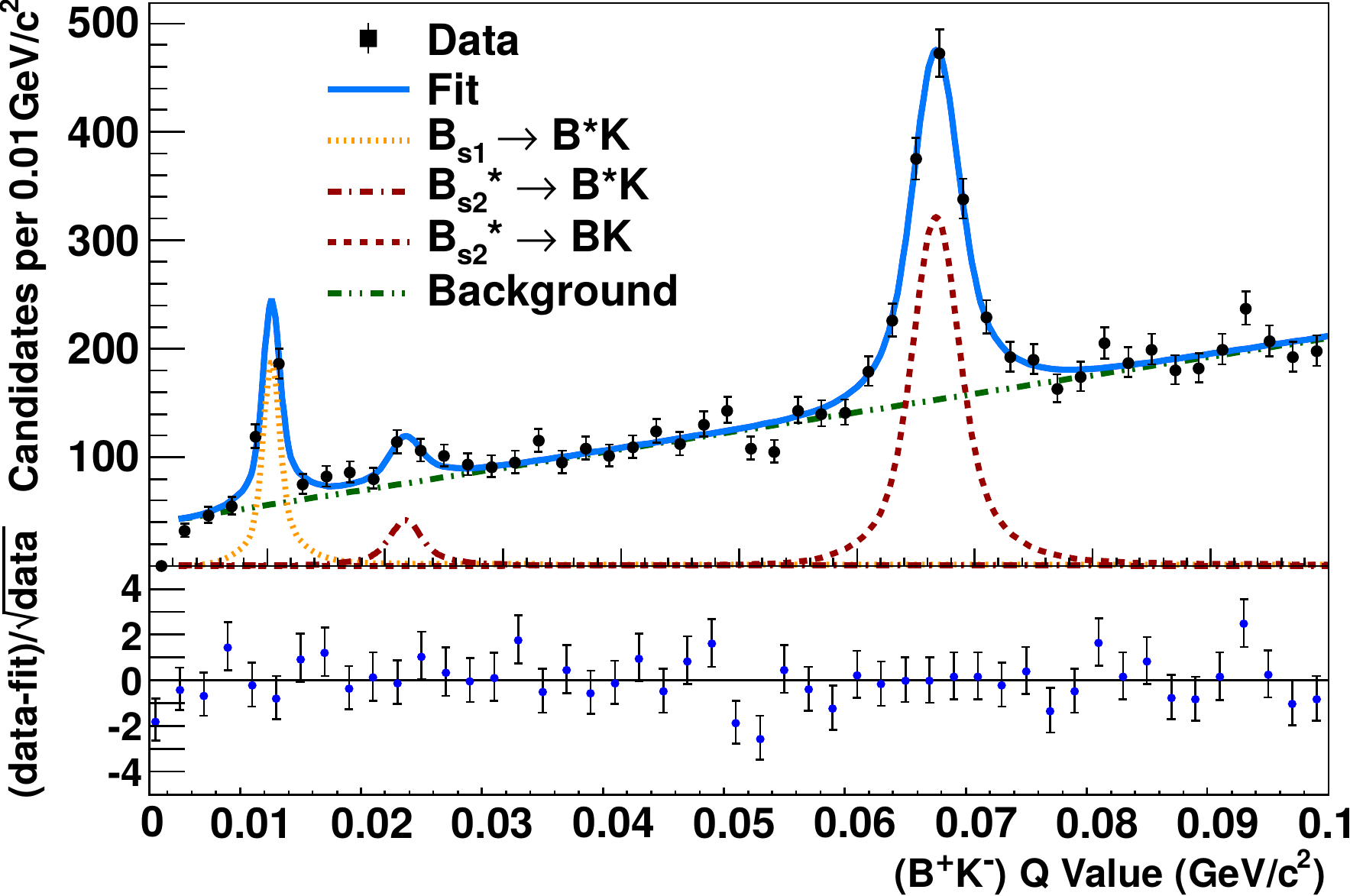} \\
  \vspace*{0.5cm}
  \includegraphics[scale=0.6]{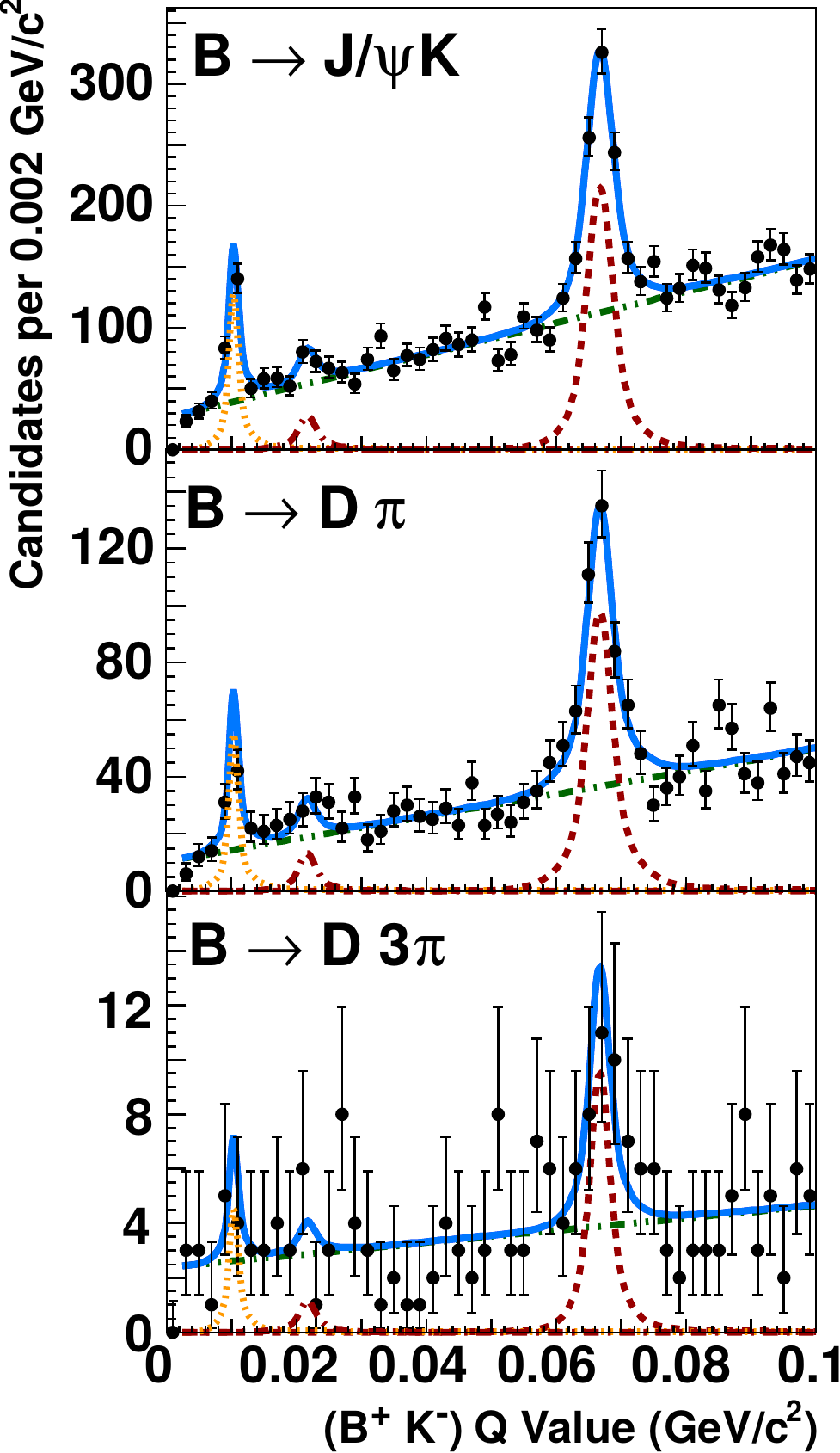}  
  \caption{Distribution of $Q$ value of $B_s^{**0}$ candidates with fit results overlaid.
The upper panel shows the data summed over decay channels and the deviations of these from the fit function,
normalized to the poisson uncertainty of the data. The lower panels show data and fits for each decay channel individually.}
	\label{fig:BsSSfit}
\end{figure}

The final selection is made by imposing a requirement on the output of the discriminator for each $B^{**}_{(s)}$ decay channel. 
The requirement is chosen by optimizing the figure of merit $N_{\rm{MC}}/\sqrt{N}$, where $N_{\rm{MC}}$ corresponds to the number of selected simulated signal events and $N$ is the number of observed events in the signal region $305 < Q < 325$ MeV/$c^{2}$ for $B^{**0}$ and $B^{**+}$ decays and $62 < Q < 72$ MeV/$c^{2}$ for $B^{**0}_s$ decays. 
For $B^{**0}$ and $B^{**+}$ candidates, the data sample is divided into a subsample with one candidate per event and a subsample with multiple candidates per event to increase sensitivity, as resulting from the better signal-to-background ratio in the single-candidate subsample.
The multiple-candidate events amount to 40-50\% of the samples. 
The resulting $B^{**}_{(s)}$-meson spectra are shown in Figs.~\ref{fig:B0SSfit} to \ref{fig:BsSSfit}.


As in earlier measurements~\cite{CDFmea,CDFsmea}, the narrow state at the lowest $Q$ value is interpreted as the $B_1 \rightarrow B^* h$ signal and the two higher $Q$-value structures as $B_2^* \rightarrow B^* h$ and $B_2^* \rightarrow B h$ signals.
In the $B^{**0,+}$ spectrum, the two lower $Q$-value signals overlap.
At $Q$ values around $550$~MeV/$c^2$ a broad structure is visible, in both the $B^{**0}$ and $B^{**+}$ invariant-mass distributions.

\section{$Q$-value fit} \label{sec:fit}
We use a maximum-likelihood fit of the unbinned $Q$-value distributions to measure the properties of the observed structures.
Separate fits are performed for $B^{**0}$, $B^{**+}$, and $B^{**0}_s$ mesons. 
For each flavor, the spectra for several $B$-meson decay channels are fit simultaneously. 
Each $Q$-value distribution is fit with the sum of various signal components and a background component. 
The signal parameters are the same in all spectra, while individual background parameters are used in each subsample. 
For the background component we use a $\Gamma$ distribution~\cite{gamma} for the $B^{**0,+}$ spectra and a polynomial for the $B^{**0}_s$ spectra.
The order of the polynomial is two for the $ B^{+} \rightarrow J/\psi\, K^{+} $ mode and one for the $ B^{+} \rightarrow \overline{D}^{0} n \pi$ modes.

Each $B$ signal is described by a Breit-Wigner shape whose parameters are free in the fit, convoluted with
a double Gaussian that accounts for the detector resolution and whose $Q$-value dependent parameters are determined from simulation. 

For the $B(5970)$ state, we use a non-relativistic Breit-Wigner shape, because we do not know its total angular momentum. For the $B_1$ and $B_2^*$ states, we use a relativistic Breit-Wigner shape to account for phase space effects in the $D$-wave decay. 
The amplitude~\cite{PDGBW} of the decay with angular momentum $J$ is given by
$$R_J(m)=\frac{M\Gamma_J(m)}{(M^2-m^2)-i M \Gamma^{\mathrm{tot}}_J(m)},$$
where $M$ is the nominal mass of the resonance and the mass-dependent width for $J=2$ is
$$\Gamma_2(m)=\Gamma\frac{M}{m}\frac{9+3r^2 Q^2+r^4 Q^4}{9+3r^2 q^2+r^4 q^4} \left( \frac{q}{Q} \right)^5.$$
Here $\Gamma$ is the nominal width, $q$ the momentum of the decay products in the rest frame of the mother particle, and $Q=q(M)$.
For the radius parameter, we assume a value of $r=3.5\ $GeV$^{-1}$ and vary it to obtain the associated systematic uncertainty.
$\Gamma^{\mathrm{tot}}_J(m)$ is the sum over all partial widths of the mother particle. For the $B_1$ state $\Gamma^{\mathrm{tot}}_J(m) =\Gamma_J(m)$, while for the $B_2^*$ state $\Gamma^{\mathrm{tot}}_J(m)=\Gamma_J ^{B_2^* \rightarrow B^* h}(m) + \Gamma_J ^{B_2^* \rightarrow B h}(m)$.

As the most probable production process is via $S$-wave, because higher angular momenta are suppressed, the signal shape $A(m)$ is described by
$$A(m)=|R_0(m) \cdot R_2(m)|$$
with $\Gamma_0(m)= \Gamma$.
The signal model assumes no interference with the broad $B^{**}$ states.

In order to determine directly the relative rates, the relative efficiencies for reconstructing the various $B^{**}_{(s)}$ states, determined from simulation, are included in the fit model.
The relative normalization of the $B$ decay channels is free in the fit.
Because the description of the data in terms of the known contributions and a smooth background is unsatisfactory in the $500<Q<600$~MeV/$c^2$ range of the spectrum,
we introduce an additional broad structure whose model is a non-relativistic Breit-Wigner function convoluted with a single Gaussian.
The yield of the broad structure is measured relative to the $B_2^* \rightarrow B\pi$ yield.

As in previous measurements \cite{CDFmea}, external inputs from independent experimental measurements and theoretical assumptions are used in the fit to resolve the ambiguity due to the overlapping $B^{**0,+}$ signal structures.
The difference between the mean mass values of the $B_2^* \rightarrow Bh$ and $B_2^* \rightarrow B^*h$ signal structures is
constrained to the value of $ m_{B^{*+}}-m_{B^{+}} = 45.01 \pm 0.30 \pm 0.23~\textrm{MeV/}c^{2}$ for $B^+$
mesons~\cite{PDBook} and to the flavor-averaged value of $ m_{B^{*} }-m_{ B} =45.8 \pm 1.5~\textrm{MeV/}c^{2}$ in the
case for $B^0$ mesons, where the limit 
$ \left| (m_{B^{*+} }-m_{ B^{+}}) -( m_{B^{*0} }-m_{ B^{0}}) \right| < 6~\textrm{MeV/}c^{2}$ at 95\% C.L. is used to estimate the uncertainty.

In the $B^{**0,+}$ fits, the relative branching fraction $ \mathcal{B}(B_2^* \rightarrow B\pi) / \mathcal{B}(B_2^* \rightarrow B^*\pi) = 1.02 \pm 0.24 $ is used.
This is derived from the corresponding value in $D$-meson decays, $ \mathcal{B}(D_2^* \rightarrow D\pi)/\mathcal{B}(D_2^* \rightarrow D^*\pi) = 1.56 \pm 0.16 $, by taking into account the difference in
phase space and the properties of the $D$-wave decay \cite{PDBook}.
The relative branching fraction is expressed as $ \mathcal{B}(B_2^* \rightarrow B\pi)/\mathcal{B}(B_2^* \rightarrow B^*\pi) = F_b \left(k_B / k_{B^*}\right)^5$, where $k_X$ is
the momentum of the pion in the rest frame of the particle $X$ and $F_b$ is the ratio of the form factors for the two decays. 
Due to heavy quark symmetry, the relation $F_b=F_c$ is assumed, where a calculation with a Blatt-Weisskopf form factor with a radius parameter of $r=3.5$~GeV$^{-1}$ \cite{CDFwick2} is used to estimate the uncertainty of this relation.

In the $B^{**0}$ fit, a component for misreconstructed $B_s^{**0}$ mesons in which the low-energy kaon from the $B_s^{**0}$ decay is reconstructed as a pion is added. 
The shape is determined from simulation. The yield is determined as the product of the probability for $B^{**0}_s$ mesons to meet the $B^{**0}$ selection criteria, determined from simulations, times the $B_s^{**0}$ yield observed in data. 
The misreconstruction of the pion from the $B^{**0}$ decay as a kaon leads to $Q$ values above the range considered for $B_s^{**0}$ candidates.

The results of the fits are listed in Tables~\ref{tab:ResBSS} and \ref{tab:CorBSS} and shown in Figs.~\ref{fig:B0SSfit} to \ref{fig:BsSSfit}.
The correlations between fit parameters are below 20\% (30\%) for the properties of the $B(5970)^0$ ($B(5970)^+$), except for the correlation between width and yield of 81\% (76\%).

\begin{table}[h]
     \caption{Results of the simultaneous fits to the $Q$-value spectra. Uncertainties include the statistical contribution only.}
     \label{tab:ResBSS}
     \centering 
    \begin{tabular}{lllrclrclrcl}
      \hline \hline
       & &  & \multicolumn{3}{c}{$B_{1}$} &  \multicolumn{3}{c}{$B_{2}^*$} &  \multicolumn{3}{c}{$B(5970)$} \\
      \hline
      $B^{**0}$ & Yield &  & 5300 &$\pm$& 900 & 5500 &$\pm$& 500 & 2600 &$\pm$& 700 \\
       & $Q$ (MeV/$c^{2}$) &  & 262.7 &$\pm$& 0.9 & 317.9 &$\pm$& 1.2 & 558 &$\pm$& 5  \\
       & $\Gamma$ (MeV/$c^{2}$) &  & 23 &$\pm$& 3  & 22 & $^{+}_{-}$ &  $^{3}_{2}$  & 65 &$\pm$& 18  \\
       & $r_\mathrm{prod}$ &  & \multicolumn{6}{c}{$1.0\ ^{+0.2}_{-0.4}$}  &  \\
      \hline
      $B^{**+}$ & Yield & &  4100 &$\pm$& 900 & 1700 &$\pm$& 200 & 1400 &$\pm$& 500 \\
       & $Q$ (MeV/$c^{2}$) & &  262 &$\pm$& 3  & 317.7 &$\pm$& 1.2  & 541 &$\pm$& 5  \\
       & $\Gamma$ (MeV/$c^{2}$) & &  47 &$^{+}_{-}$ &  $^{12}_{10}$  & 11 &$^{+}_{-}$ &  $^{4}_{3}$  & 50 &$\pm$& 20  \\
       & $r_\mathrm{prod}$ & &  \multicolumn{6}{c}{$2.5\ ^{+1.6}_{-1.0}$}  &  \\
      \hline
      $B_s^{**0}$ & Yield & &  280 &$\pm$& 40 & 1110 &$\pm$& 60 & \\
       & $Q$ (MeV/$c^{2}$) & &  10.35 &$\pm$& 0.10 \  & 66.73 &$\pm$& 0.13  \\
       & $\Gamma$ (MeV/$c^{2}$) & &  0.5 &$\pm$& 0.3  & 1.4 &$\pm$& 0.4  \\
       & $r_\mathrm{prod}$ & &  \multicolumn{6}{c}{$0.25\ ^{+0.07}_{-0.05}$}  &  \\
       & $r_\mathrm{dec}$ & &  \multicolumn{6}{c}{$0.10\ ^{+0.03}_{-0.02}$}  &  \\
      \hline \hline
     \end{tabular} \\
\end{table}

\begin{table}[h]
     \caption{Correlations between parameters of the simultaneous fits to the $Q$-value spectra.}
     \label{tab:CorBSS}
     \centering 
    \begin{tabular}{lrrrrrr}
      \hline \hline
       & & $\ \Gamma(B_1)$ & $\ Q(B_2^*)$ & $\ \Gamma(B_2^*)$ & $\ \ r_\mathrm{prod}$ & $\ \ \ r_\mathrm{dec}$ \\
      \hline
      $B^{**0}$ & $Q(B_1)$ & $0.39\ $ & $-0.22$ & $0.14$ & $0.40$ &  \\
       & $\Gamma(B_1)$ &  & $-0.32$ & $-0.25$ & $0.66$ &  \\
       & $Q(B_2^*)$ &  &  & $0.03$ & $0.06$ &  \\
       & $\Gamma(B_2^*)$ &  &  &  & $-0.47$ &  \\
      \hline
      $B^{**+}$ & $Q(B_1)$ & $0.37\ $ & $-0.03$ & $-0.12$ & $0.54$ &  \\
       & $\Gamma(B_1)$ &  & $-0.14$ & $-0.15$ & $0.32$ &  \\
       & $Q(B_2^*)$ &  &  & $-0.03$ & $0.09$ &  \\
       & $\Gamma(B_2^*)$ &  &  &  & $-0.51$ &  \\
      \hline
      $B_s^{**0}$ & $Q(B_1)$ & $-0.29\ $ & $0.03$ & $-0.01$ & $-0.25$ & $-0.01$ \\
       & $\Gamma(B_1)$ &  & $0.02$ & $0.01$ & $0.79$ & $0.03$ \\
       & $Q(B_2^*)$ &  &  & $0.08$ & $-0.03$ & $0.01$ \\
       & $\Gamma(B_2^*)$ &  &  &  & $-0.23$ & $0.28$ \\
       & $r_\mathrm{prod}$ &  &  &  &  & $-0.06$ \\

      \hline \hline
     \end{tabular} \\
\end{table}

To measure the relative rate of $B$ and $B^{**0}$ mesons production, 
we use the ratio between the sum of $B_1^0$ and $B_2^{0*}$ meson 
yields reconstructed in the $B^{**0}\rightarrow B^{+(*)}\pi^-$ decay, followed by the $B^{+} \rightarrow \overline{D}^{0} \pi^{+}$ decay, and $B^+$ meson yields reconstructed in the same final state.
The conditional probability for reconstructing a $B^{**0}$ meson if a $B^+$ meson is already reconstructed in a $B^{**0}\rightarrow B^{(*)+}\pi^-$ event is determined from simulation.
Under the assumption of isospin symmetry, $B^{**0}$ mesons decay to $B^0 \pi^0$ states in one third of the cases and are therefore not reconstructed. 
After correcting for efficiency and for the unreconstructed decays involving neutral pions, we find
that $19 \pm 2 \textrm{(stat)}\%$ of the events with a $B^+$ meson with $p_T > 5$~GeV$/c$ contain a $B^{**0}$ meson.

\section{Systematic uncertainties}
Several sources of systematic uncertainties are considered, including uncertainties on the absolute mass scale, mass resolution, and the fit model. The size of systematic uncertainties considered are listed in Tables~\ref{tab:AllSystB0} to \ref{tab:SystX}.
The study of the mass-scale uncertainty was performed in earlier $B^{**}_{(s)}$ analyses \cite{CDFmea,CDFsmea} by reconstructing $\psi(2S) \rightarrow J/\psi\,\pi^+\pi^-$ and $D^{**} \rightarrow D^{(*)+}\pi^-$ control channels and comparing the $Q$ values observed in these with the known values.

The detector resolution was studied in a previous analysis \cite{CDFwick},
using final states with similar topology and kinematic regime as in the present measurement. 
The modes investigated included $D^{*+} \rightarrow D^0\pi^+$ and $\psi(2S) \rightarrow J/\psi\, \pi^+\pi^-$ decays, with $Q$ values 6~MeV/$c^2$ and 310~MeV/$c^2$, respectively.
The method is improved for the present analysis.
First we rescale the mass resolution of the simulation to match the resolution observed in data,
using a $Q$-value-dependent factor linearly interpolated from the $Q$ values observed in the reference channels.
To estimate the systematic uncertainty of the scale factor,
we study its variation as a function of the transverse momentum of the pion from the $D^{*+}$ meson decay
and of the pion pair from the $\psi(2S)$ meson decay.
The chosen uncertainty is such that all determined scale factors are within one standard deviation.
A difference between simulation and experimental data is expected, because the simulation does not model accurately the particle multiplicity of the data. 
Additional particles present in data are expected to reduce the efficiency of associating drift-chamber hits to the tracks.
The loss of hits worsens the mass resolution by 5\% for $B^{**0,+}$ and 10\% for $B^{**0}_s$ decays,
both with an uncertainty of 5\%.

The systematic uncertainty associated with the signal model is quantified by varying the radius parameter $r$ to $0$ and $4$~GeV$^{-1}$ and taking the largest difference to the nomial fit. This effect is negligible in comparison to other sources of uncertainty.

The systematic uncertainty associated with possible mismodelings of the background shape is estimated by fitting with alternative background models
and taking the deviation of the results with respect to the default fit as the uncertainty.
For $B^{**0,+}$ mesons, the alternative fit model is
a polynomial function. For the $B^{**0}_s$ spectrum, a polynomial function one order higher than the default model is used.
Two broad $B^{**0,+}$ $j=\frac{1}{2}$ states are expected at similar masses as the two narrow $B^{**0,+}$ states, but predictions for their masses and widths vary significantly.
To assess a systematic uncertainty associated with the limited knowledge of resonance parameters of broad states, we perform 100 fits with two additional Breit-Wigner functions for these states in the fit model.
Their $Q$ values are varied between 200 and 400~MeV/$c^2$ and the widths between 100 and 200 MeV/$c^2$. 
The largest deviation in the estimate of each signal parameter with respect to the results of the default fit
is taken as systematic uncertainty.
The mass spectrum of $B^{**0}_s$ candidates is steeply rising at the kinematic threshold.
The default fit starts from 5~MeV/$c^2$ using a relatively simple background shape.
The lower bound of the fit is varied by $\pm5$~MeV/$c^2$ and the largest difference in fit results with respect to the default fit is taken as an additional uncertainty on the background model.

To test for biases in the fitting procedure, we simulate random mass spectra with known signal parameters and fit them with the default model.
Some of the fit parameter estimates show mild biases, which never exceed 30\% of the statistical uncertainty.
The estimates showing nonzero biases are corrected for their bias and the full size of the bias is assigned as a systematic uncertainty.
The assumed photon energy from the $B^*$ decay and the branching fraction of the $B_2^*$ decays are varied within their uncertainties and the data are fit again.
The deviations in the measured parameters with respect to the default results are taken as systematic uncertainties.
In the $B^{**0}$ and $B^{**+}$ fits, these uncertainties are usually positively correlated, except for the uncertainty on $\Gamma(B_2^*)$, where an anti-correlation is observed.

The relative acceptance between $B_{(s)1} \rightarrow B^{*} h$, $B_{(s)2}^{*} \rightarrow B^{*} h$, and $B_{s2}^{*} \rightarrow B h$ decays derived from simulation varies between 0.9 and 1.1 for $B^{**0,+}$ mesons and between 0.95 and 1.05 for $B^{**0}_s$ mesons.
We assign a relative uncertainty of 10\% and 5\%, respectively, on the measurement of the relative branching fractions.
The influence of the non-flat relative acceptance on the measurement of the signal properties is estimated with pseudoexperiments where a modified acceptance is applied to the generated signal mass spectra.

\begin{table}[h]
	\caption{Systematic and statistical uncertainties in the $B^{**0}$ measurements.}
    \label{tab:AllSystB0}
\centering
\begin{tabular}{lcccccc}
\hline  \hline
 & \multicolumn{2}{c}{$Q$ (MeV/$c^{2}$)} & \multicolumn{2}{c}{$\Gamma$ (MeV/$c^{2}$)} & $\Delta m$ & $r_\mathrm{prod}$ \\ 
 & $B_1$ & $B_2^*$ & $B_1$ & $B_2^*$ & (MeV/$c^{2}$) & \\ 
\hline 
Mass scale 				& \ws 0.2 & \ws 0.2 & \ws -   & \ws - & \ws 0.0 & \ws -    \\ 
Resolution 				& \ws 0.0 & \ws 0.0 & \ws 0.3 & \ws 0.2 & \ws 0.0 & \ws 0.0 \\ 
Signal Model 			& \ws 0.0 & \ws 0.1 & \ws 0.7 & \ws 0.7 & \ws 0.1 & \ws 0.1 \\ 
Backgr. model 			& \ws 0.0 & \ws 0.7 & \ws 3.2 & \ws 3.6 & \ws 0.7 & \ws 0.3 \\ 
Broad $B^{**0}$ states	& \ws $^{+0.1}_{-0.3}$  & \ws $^{+0.0}_{-0.4}$ & \ws $^{+0.1}_{-2.1}$ & \ws $^{+0.0}_{-3.9}$  & \ws $^{+0.3}_{-0.4}$ & \ws $^{+0.3}_{-0.0}$  \\ 
Fit bias 				& \ws -  & \ws -  & \ws -  & \ws 0.3 & \ws - & \ws $^{+0.0}_{-0.1}$ \\ 
Fit constraints 		& \ws 1.1 & \ws $^{+0.3}_{-0.2}$ & \ws $^{+1.5}_{-1.6}$ & \ws 0.4 & \ws 0.9 & \ws $^{+0.2}_{-0.3}$ \\ 
Acceptance 				& \ws -   & \ws $^{+0.0}_{-0.3}$   & \ws $^{+0.6}_{-0.0}$ & \ws - & \ws $^{+0.3}_{-0.0}$   & \ws $^{+0.1}_{-0.2}$ \\ 
\hline 
Total systematic		& \ws $^{+1.1}_{-1.2}$ & \ws $^{+0.8}_{-0.9}$ & \ws 4 & \ws $^{+4}_{-5}$ & \ws 1.2 & \ws 0.5 \\ 
\hline 
Statistical 			& \ws 0.9 & \ws 1.2 & \ws 3 & \ws $^{+3}_{-2}$ & \ws 1.7 & \ws $^{+0.2}_{-0.4}$ \\     
\hline \hline 
\end{tabular}
\end{table}

\begin{table}[h]
    \caption{Systematic and statistical uncertainties in the $B^{**+}$ measurements.}
    \label{tab:AllSystBp}
\centering
\begin{tabular}{lcccccc}
\hline \hline 
 & \multicolumn{2}{c}{$Q$ (MeV/$c^{2}$)} & \multicolumn{2}{c}{$\Gamma$ (MeV/$c^{2}$)} & $\Delta m$ & $r_\mathrm{prod}$ \\ 
 & $B_1$ & $B_2^*$ & $B_1$ & $B_2^*$ & (MeV/$c^{2}$) & \\ 
\hline 
Mass scale 				& \ws 0.2 & \ws 0.2 & \ws -   & \ws - &   \ws 0.0 & - \\ 
Resolution 				& \ws 0.0 & \ws 0.0 & \ws 0.1 & \ws 0.2 & \ws 0.0 & \ws 0.0 \\ 
Signal Model 		 	& \ws 0.3 & \ws 0.0 & \ws 1.0 & \ws 0.7 & \ws 0.3 & \ws 0.2 \\ 
Backgr. model 			& \ws 0.4 & \ws 0.1 & \ws 2.1 & \ws 1.6 & \ws 0.5 & \ws 0.5 \\ 
Broad $B^{**+}$ states	& \ws $^{+0.2}_{-1.3}$ & \ws $^{+0.1}_{-0.0}$ & \ws $^{+0.0}_{-9.9}$  & \ws $^{+0.0}_{-3.2}$ & \ws $^{+1.3}_{-0.0}$ & \ws $^{+0.5}_{-0.2}$ \\ 
Fit bias 				& \ws - & \ws - & \ws $^{+0.0}_{-1.9}$  & \ws - & \ws - & \ws $^{+0.0}_{-0.4}$ \\ 
Fit constraints 		& \ws $^{+1.0}_{-2.2}$ & \ws $^{+0.2}_{-0.9}$ & \ws $^{+0.0}_{-7.4}$  & \ws $^{+2.8}_{-1.4}$ & \ws $^{+1.5}_{-0.8}$ & \ws $^{+0.5}_{-0.8}$ \\ 
Acceptance 				& \ws - & \ws 0.0   & \ws $^{+0.0}_{-3.8}$  & \ws $^{+1.0}_{-0.0}$ & \ws 0.0 & \ws $^{+0.3}_{-0.5}$ \\ 
\hline 
Total systematic 		& \ws $^{+1}_{-3}$ & \ws $^{+0.3}_{-0.9}$ & \ws $^{+2}_{-13}$ & \ws $^{+3}_{-4}$ & \ws $^{+2}_{-1}$ & \ws $^{+0.9}_{-1.2}$ \\ 
\hline 
Statistical 			& \ws 3 & \ws 1.2 & \ws $^{+12}_{-10}$ & \ws $^{+4}_{-3}$  & \ws 3 & \ws $^{+1.6}_{-1.0}$ \\ 
\hline \hline 
\end{tabular} 
\end{table}

\begin{table}[h]
	\caption{Systematic and statistical uncertainties in the $B_s^{**0}$ measurements.}
    \label{tab:AllSystBs}
\centering
\begin{tabular}{lccccccc}
\hline  \hline 
 & \multicolumn{2}{c}{$Q$ (MeV/$c^{2}$)} & \multicolumn{2}{c}{$\Gamma$ (MeV/$c^{2}$)} & $\Delta m$ & $r_\mathrm{prod}$ & $r_\mathrm{dec}$ \\ 
 & $B_{s1}$ & $B_{s2}^*$ & $B_{s1}$ & $B_{s2}^*$ & (MeV/$c^{2}$) & & \\ 
\hline 
Mass scale 		& \ws 0.14 & \ws 0.14 & \ws -    & \ws -    & \ws 0.01 & \ws -    & \ws -    \\ 
Resolution 		& \ws 0.00 & \ws 0.00 & \ws 0.06 & \ws 0.19 & \ws 0.00 & \ws 0.00 & \ws 0.00 \\ 
Signal Model 	& \ws 0.00 & \ws 0.00 & \ws 0.00 & \ws 0.00 & \ws 0.00 & \ws 0.00 & \ws 0.00 \\ 
Bkg. model 		& \ws 0.00 & \ws 0.01 & \ws 0.02 & \ws 0.05 & \ws 0.01 & \ws 0.00 & \ws 0.01 \\ 
Fit range 		& \ws 0.04 & \ws 0.01 & \ws 0.26 & \ws 0.02 & \ws 0.03 & \ws 0.04 & \ws 0.01 \\ 
Fit bias 		& \ws -    & \ws -    & \ws 0.02 & \ws 0.02 & \ws -    & \ws $^{+0.00}_{-0.01}$    & \ws -    \\ 
Fit constr.		& \ws 0.00 & \ws 0.02 & \ws 0.01 & \ws 0.04 & \ws 0.02 & \ws 0.00 & \ws 0.00 \\ 
Acceptance 		& \ws -    & \ws -    & \ws -    & \ws -    & \ws -    & \ws 0.01 & \ws 0.01 \\ 
\hline 
Total syst. 	& \ws 0.15 & \ws 0.14 & \ws 0.3 & \ws 0.2 & \ws 0.04 & \ws 0.04 & \ws 0.02 \\ 
\hline 
Statistical 	& \ws 0.12 & \ws 0.13 & \ws 0.3 & \ws 0.4 & \ws 0.18 & \ws $^{+0.07}_{-0.05}$ & \ws $^{+0.03}_{-0.02}$ \\ 
\hline \hline 
\end{tabular}    
\end{table}

\begin{table}[h]
    \caption{Systematic and statistical uncertainties in the neutral and charged $B(5970)$ measurements.}
    \label{tab:SystX}
  \centering
    \begin{tabular}{lcccccc}
      \hline \hline
         & \multicolumn{2}{c}{$Q$ (MeV/$c^{2}$)} & \multicolumn{2}{c}{$\Gamma$ (MeV/$c^{2}$)} & \multicolumn{2}{c}{Rel. yield} \\ 
         & Neutr. & Char. & Neutr. & Char. & Neutr. & Char. \\ 
      \hline
	Bkg. model 		& 12 & 12 & 30 & 40 & 0.3 & 0.8 \\ 
	Fit bias 		& - & - & $^{+0}_{-10}$ & $^{+0}_{-10}$ & - & - \\ 
	Acceptance 		& - & $^{+1}_{-0}$ & 3 & - &  $^{+0.2}_{-0.1}$ & $^{+0.2}_{-0.1}$ \\       
      \hline  \hline
	Total syst. 	& 12 & 12 & 30 & 40 & $^{+0.4}_{-0.3}$ & 0.8 \\ 
	\hline 
	Statistical  	& $\ $5 & 5 & $^{+30}_{-20}$ & $^{+30}_{-20}$ & $^{+0.2}_{-0.1}$ & $^{+0.3}_{-0.2}$ \\ 
      \hline  \hline
    \end{tabular} \\    
\end{table}

The conditional probability for reconstructing a $B^{**0}$ meson if a $B^+$ meson is already reconstructed
depends on the transverse momentum of the $B^{**0}$ mesons.
The $B^{**0}$-meson yields in data and simulated events are compared in six independent ranges of transverse momentum.
As they are found to be consistent, no correction is applied.
To estimate a systematic uncertainty on the efficiency, the ratio of yields is fit with a straight line,
which is used to weight the generated spectrum in the simulations.
The resulting 20 \% change in efficiency is taken as the systematic uncertainty of the relative rate of $B$ and $B^{**0}$ mesons production.

The dominant systematic uncertainty for most quantities is the description of the background shape, except for the $Q$ values of the $B^{**0}_s$ states, where the mass-scale uncertainty dominates.
For the $B^{**0,+}$ states an additional significant contribution comes from the fit constraints.
Because the $B_2^* \rightarrow B\pi$ signal is well separated from the overlapping signals, the $B_2^*$ properties are less affected by this systematic uncertainty.

\section{Evidence for a $B(5970)$ state}
\begin{figure}[h]
  \centering  
  \includegraphics[width=0.47\textwidth]{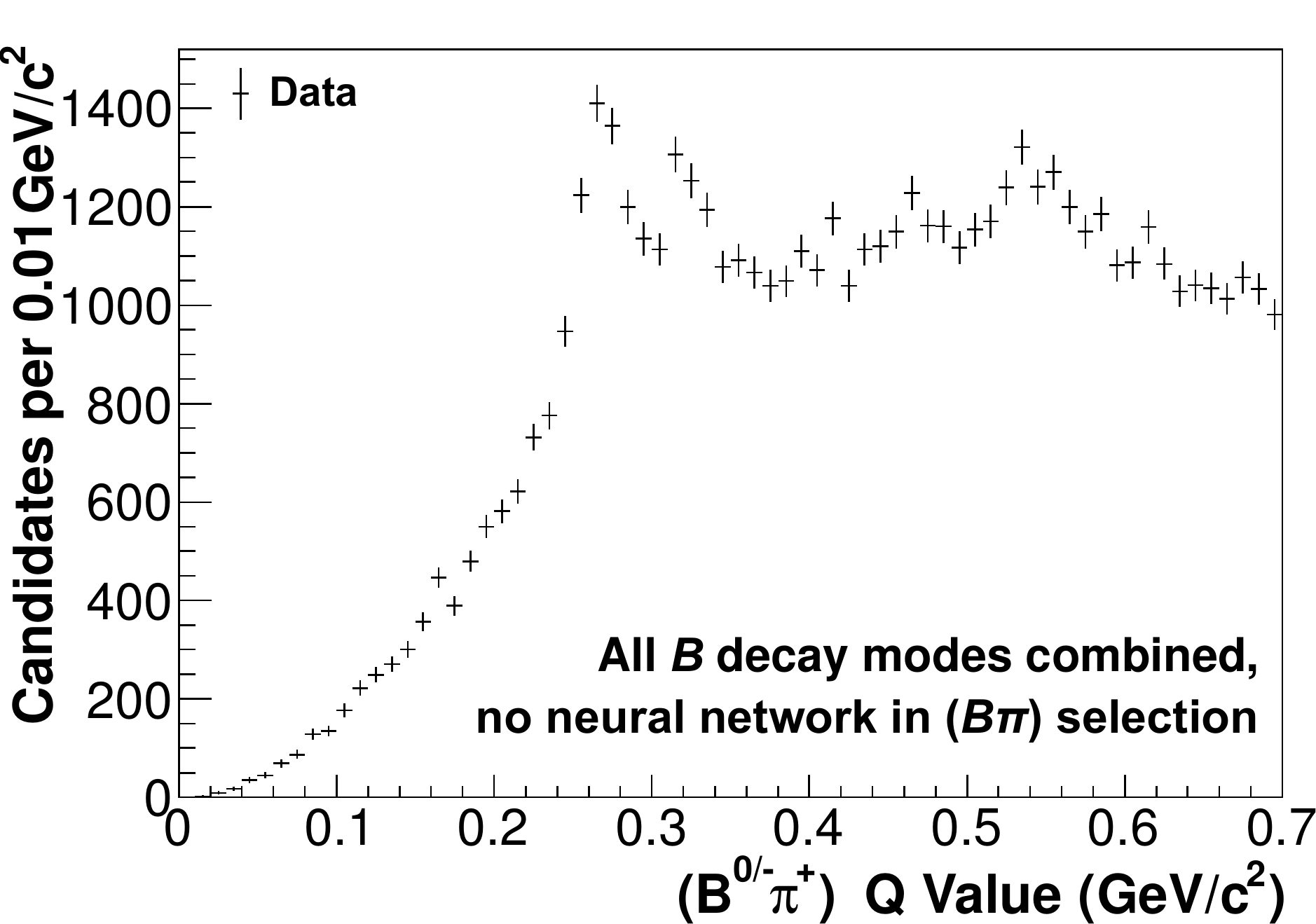}
  \caption{$Q$-value distribution of $B^-\pi^+$ and $B^0\pi^+$ candidates selected with alternative requirements.}
	\label{fig:BSScut}
\end{figure}

As a consistency check that the structure at $Q \approx 550$~MeV/$c^2$ is not an artifact of the selection, we apply to $B^+\pi^+$ combinations the same criteria as for the signal sample.
No structure is observed in the invariant-mass distribution of the wrong-charge combinations as shown in Fig.~\ref{fig:B0SSfit}.
Because $B^0$ mesons oscillate this cross check cannot be done with $\bar{B}^0\pi^+$ combinations.
The new signal is verified to be robust against significant variations of the selection requirements, as shown in Fig.~\ref{fig:BSScut}, where a requirement on the transverse momentum of the pion instead of a requirement on the output of the neural network is applied.
As we have no sensitivity to determine whether the enhancement is caused by multiple overlapping broad states or not, we treat it as a single resonance in the following.

\begin{figure}[h]
  \centering  
  \includegraphics[width=0.43\textwidth]{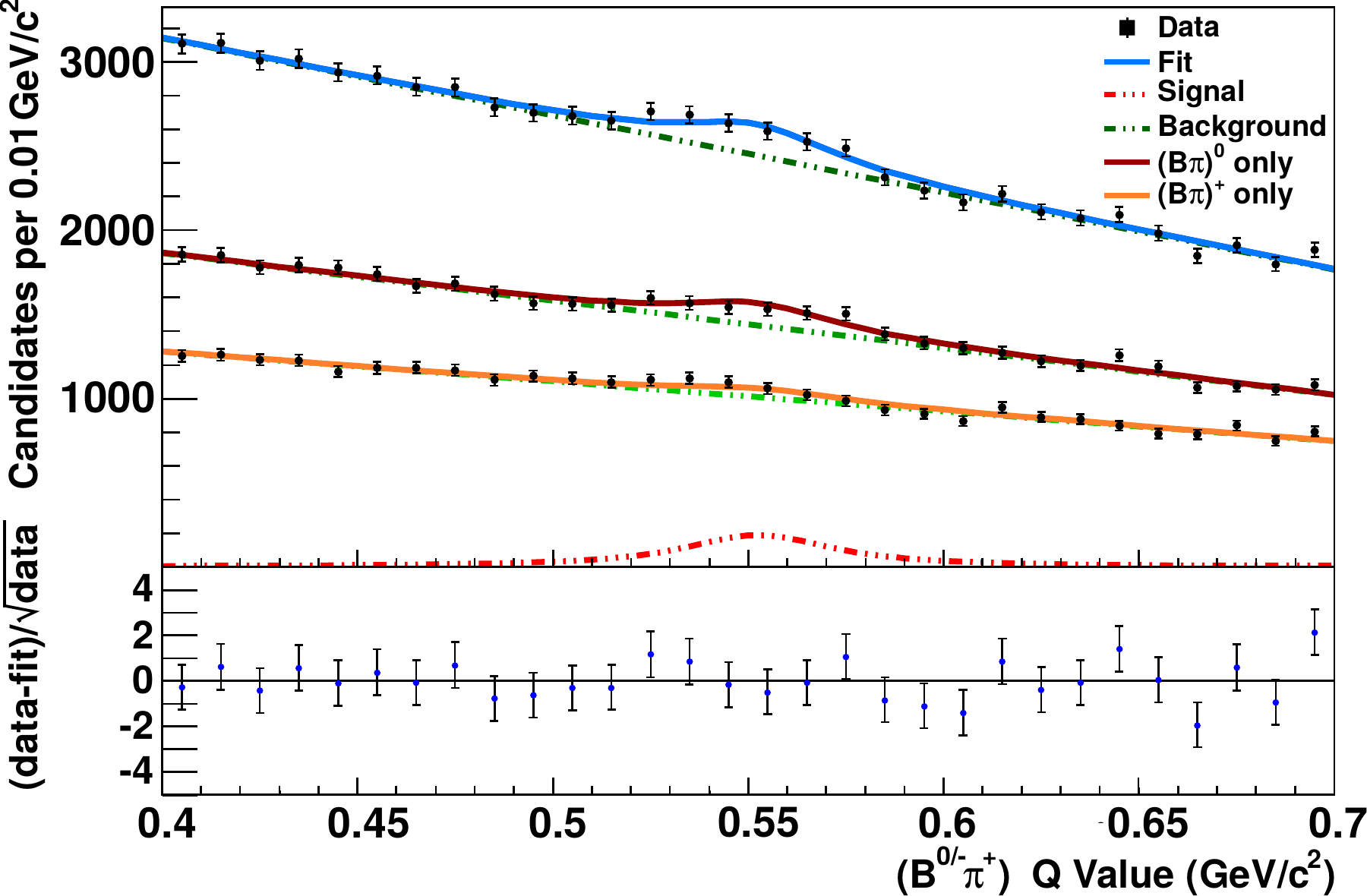} \\
  \vspace*{0.5cm}
  \includegraphics[width=0.33\textwidth]{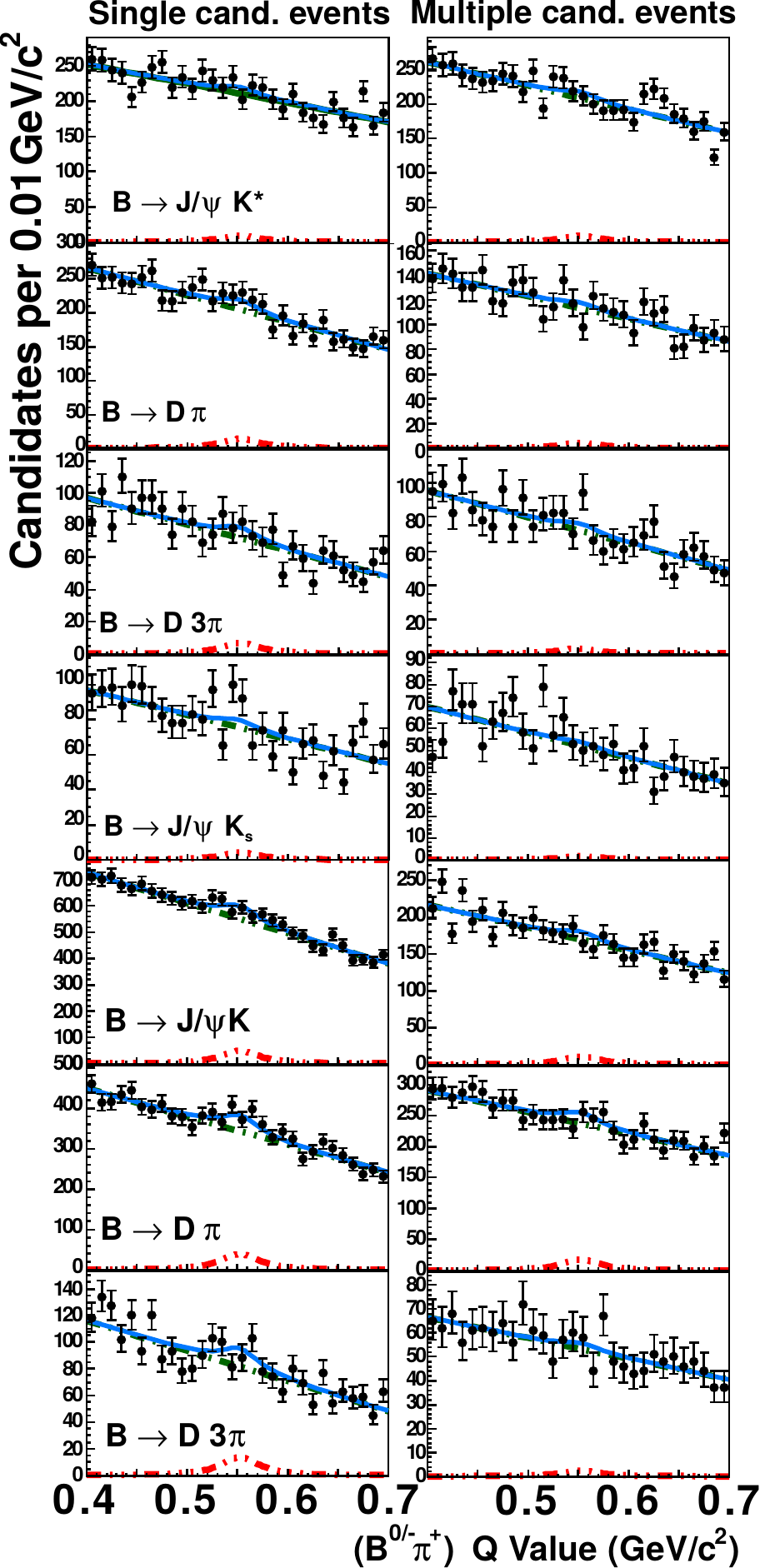}  
  \caption{Spectra of $Q$ value of $B^{**0,+}$ candidates in all the considered decay channels with fit results for the broad structure overlaid. 
The upper panel shows the data summed over decay channels and the deviations of these from the fit function,
normalized to the poisson uncertainty of the data.
The lower plot shows the simultaneously-fit spectra separately.}
	\label{fig:Xfit}
\end{figure}

To determine the significance of the previously unobserved broad structure, we use the difference $\Delta L$
in logarithms of the likelihood between data fits that include or not the $B(5970)^{0,+}$ signal component.
The $B(5970)^{0}$ and $B(5970)^{+}$ candidates are fit simultaneously with common signal parameters.
Using random distributions generated from the background distribution observed in the data, we determine the probability $p$ of observing a value of $\Delta L$ at least as large as that observed in data. 
We restrict the fit range to $Q>400$~MeV/$c^2$ because at lower values a broad structure would be indistinguishable from the background of the $B^{**0,+}$ states. 
In the range studied, the background is described by a straight line.
In the fits that allow for the presence of a $B(5970)^{0,+}$ component, the signal yield
is floating freely, and the mean and width are constrained to be in the ranges 450 to 650 MeV/$c^2$ and 10 and 100 MeV/$c^2$, respectively,
to avoid having a large fraction of the signal outside the fit range.
The result of the fit to data is shown in Fig.~\ref{fig:Xfit}.
We observe a $\Delta L$ value of 18 in data.
A higher value is obtained in only 128 of $1.2 \times 10^7$ background-only pseudoexperiments, corresponding to a statistical significance of $4.4 \sigma$.

To check the systematic effect of the background model on the significance, we repeat the significance evaluation with the default fit model of the $B^{**0,+}$ measurement, but with fixed $B^{**0,+}$ signal parameters.
Independent parameters are used for the $B(5970)^{0}$ and $B(5970)^{+}$ signals, where we find individual significances of $4.2 \sigma$ and $3.7 \sigma$ for the neutral and charged state respectively.
In combination, with the alternative fit model we obtain a significance higher than with the default fit. 

\section{Results}
We measure the masses and widths of fully reconstructed $B^{**0}$, $B^{**+}$, and $B^{**0}_s$ mesons.
The sample contains approximately 10800 $B^{**0}$ decays, 5800 $B^{**+}$ decays, and 1390 $B^{**0}_s$ decays.
The results are shown in Table~\ref{tab:ResBSS2}.
In addition, the relative production rates of $B_{1}$ and $B_{2}^{*}$ multiplied by their branching fraction into the analyzed decay channels are measured and their values are listed in Table~\ref{tab:Resrprod}.
The determination of the relative branching fractions of the $B_{s2}^{*}$ state as defined in Eq.(\ref{eqn:rdec}) yields
$ r_\mathrm{dec} = 0.10 \ ^{+0.03}_{-0.02}\ (\textrm{stat}) \pm 0.02\ (\textrm{syst})$.

\begin{table}[h]
     \caption{Measured masses and widths of $B^{**_{(s)}}$ mesons. The first contribution to the uncertainties is statistical; the second is systematic.}
     \label{tab:ResBSS2}
	\centering
    \begin{tabular}{lccc}
      \hline \hline
       & $Q$ (MeV$/c^2$) & & $\Gamma$ (MeV$/c^2$)\\
      \hline
      $B_{1}^{0}$   & $262.7 \pm 0.9 \  ^{+1.1}_{-1.2}$ 	& & $23 \pm 3 \pm 4$ \\
      $B_{2}^{*0}$  & $317.9 \pm 1.2  \ ^{+0.8}_{-0.9}$ 	& & $ 22 \ ^{+\ 3}_{-\ 2} \ ^{+\ 4}_{-\ 5} $ \\
      $B_{1}^{+}$   & $262\ \ \,  \pm 3\ \ \,   \ ^{+1}_{-3}\ \ \,  $ 	& & $49 \ ^{+12}_{-10}  \ ^{+\ \,2}_{-13}$ \\
      $B_{2}^{*+}$  & $317.7 \pm 1.2  \ ^{+0.3}_{-0.9}$ 	& & $11 \ ^{+\ 4}_{-\ 3} \ ^{+\ 3}_{-\ 4} $ \\
      $B_{s1}^{0}$  & $10.35 \pm 0.12 \pm 0.15$ 	& & $0.5 \pm 0.3 \pm 0.3$ \\
      $B_{s2}^{*0}$ & $66.73 \pm 0.13 \pm 0.14$ 	& & $1.4 \pm 0.4 \pm 0.2$ \\
      \hline \hline
     \end{tabular} \\
\end{table}

\begin{table}[h]
     \caption{Measured $B^{**}_{(s)}$ meson relative production rates times branching fractions as defined in Eq.(\ref{eqn:rprod}) in the visible range $p_T>5$~GeV$/c$. The first contribution to the uncertainties is statistical; the second is systematic.
}
     \label{tab:Resrprod}
	\centering
    \begin{tabular}{lc}
      \hline \hline
       & $r_\mathrm{prod}$ \\
      \hline
      $B^{**0}$  & $1.0 \ \,  ^{+0.2}_{-0.4} \ \pm 0.5\ \, $ \\
      $B^{**+}$  & $2.7 \ \,  ^{+1.6}_{-1.0}\ \  ^{+0.9}_{-1.2}\ \ \ $ \\
      $B_s^{**}$ & $0.25  ^{+0.07}_{-0.04} \pm 0.05$ \\
      \hline \hline
     \end{tabular} \\
\end{table}

We also determine how many narrow $B^{**0}$ states, $B_1^0$ and $B_2^{*0}$, are produced per $B^+$ meson.
For $B^+$ mesons having a transverse momentum larger than 5~GeV$/c$ the fraction is $19\pm 2(\textrm{stat})\pm 4(\textrm{syst})\%$.

The properties of the previously unobserved resonance are measured for neutral and charged states separately in a sample that contains 2600 $B(5970)^0$ and 1400 $B(5970)^+$ decays as shown in Table~\ref{tab:ResX}.
Assuming a decay through the $B \pi$ channel, we calculate the masses $m(B(5970)^{0}) = 5978 \pm 5 \pm 12$~MeV$/c^2$ and $m(B(5970)^{+}) = 5961 \pm 5 \pm 12$~MeV$/c^2$.
For a decay to the $B^* \pi$ final state the masses would increase by $m_{B^* }-m_{ B}$.

Assuming heavy-quark symmetry, we compare these results to the corresponding values observed for excited $D$ mesons. 
States at higher masses than $D^{**}$ excitations have been observed \cite{PDBook}.
The $D(2750)$ meson has a natural width of $63\pm6$~MeV$/c^2$ and a mass about 750~MeV$/c^2$ higher than the $D^*$ mass.
An analoge excitation of the $B^*$ would have a mass of about 6075~MeV$/c^2$ and the partner of the $B$ ground state would be expected at approximately $6030$~MeV$/c^2$.
A decay to a $B^*\pi$ state but not to a $B\pi$ state due to angular momentum and parity could lead to a reconstructed invariant mass of approximately 5985~MeV$/c^2$.

In Ref.~\cite{PredM} the only predicted states with mass values between the $B^{**0,+}$ masses and 
6100~MeV/$c^2$ are the two radial excitations $2(^1S_0)$ and $2(^3S_1)$,
with masses of 5890 and 5906~MeV/$c^2$, respectively. 
The next orbital $B$ excitation, expected to decay by $D$-wave having $L=2$, is at a mass near 6100~MeV$/c^2$.

\begin{table}[h]
     \caption{Observed resonance parameters of the broad structures. The first contribution to the uncertainties is statistical; the second is systematic.}
     \label{tab:ResX}
\centering
    \begin{tabular}{lccc}
      \hline \hline
       & $Q$ (MeV$/c^2$) & & $\Gamma$ (MeV$/c^2$)\\
      \hline
      $B(5970)^{0}$ & $558 \pm 5 \pm 12$ & & $70 \, ^{+30}_{-20} \pm 30$ \\
      $B(5970)^{+}$ & $541 \pm 5 \pm 12$ & & $60 \, ^{+30}_{-20} \pm 40$ \\
      \hline \hline
     \end{tabular} \\
\end{table}

We measure the rates of the broad structures relative to the decays $B_2^* \rightarrow B \pi$ in the range $p_T > 5$~GeV$/c$ of the produced $B$ meson,
\begin{equation}
r_\mathrm{prod}^\prime(B(5970)) = \frac{\sigma (B(5970))}{\sigma (B_{2}^{*})}  \frac{\mathcal{B}(B(5970) \rightarrow B^{(*)+} \pi^{-})}
		{ \mathcal{B}(B_{2}^{*} \rightarrow B \pi)},
\label{eqn:rprodprime}
\end{equation}
to be
$r_\mathrm{prod}^\prime(B(5970)^0) = 0.5 \ ^{+0.2}_{-0.1}\ (\textrm{stat}) \ ^{+0.4}_{-0.3}\ (\textrm{syst}) $
and
$r_\mathrm{prod}^\prime(B(5970)^+) = 0.7 \ ^{+0.3}_{-0.2}\ (\textrm{stat}) \pm 0.8\ (\textrm{syst}) $.

We calculate the masses of all states from the measured $Q$ values using known values \cite{PDBook} for the pion, kaon, and $B$-meson masses and $m_{ B^{*0,+}}-m_{B^{0,+} }$. For the $B(5970)$ state we assume the decay to $B\pi$. The results are shown in Table~\ref{tab:masses}.

\begin{table}[h]
     \caption{Masses of the observed states. The first contribution to the uncertainties is statistical; the
second is systematic; the third is the uncertainty on the known values for the $B$-meson masses and for the mass difference $m_{ B^{*0,+}}-m_{B^{0,+} }$.}
     \label{tab:masses}
\centering
    \begin{tabular}{lc}
      \hline \hline
       & $m$ (MeV$/c^2$) \\
      \hline
      $B_{1}^{0}$ 		& $5726.6 \pm 0.9  \ ^{+1.1}_{-1.2} \pm 0.4$  \\
      $B_{2}^{*0}$ 		& $5736.7 \pm 1.2  \ ^{+0.8}_{-0.9} \pm 0.2$  \\
      $B_{1}^{+}$ 		& $5727\ \,\,  \pm 3\ \,\   \ ^{+1}_{-3}\ \,\  \pm 2\ \,\, $    \\
      $B_{2}^{*+}$ 		& $5736.9 \pm 1.2  \ ^{+0.3}_{-0.9} \pm 0.2$  \\
      $B_{s1}^0$ 		& $5828.3 \pm 0.1  \pm 0.2 \pm 0.4$  \\
      $B_{s2}^{*0}$  	& $5839.7 \pm 0.1  \pm 0.1 \pm 0.2$  \\
      $B(5970)^0$ 		& $5978   \pm 5    \pm 12$           \\
      $B(5970)^+$ 		& $5961   \pm 5    \pm 12$            \\  
      \hline \hline
     \end{tabular} \\
\vspace{0.3cm} 
\centering
    \begin{tabular}{lc}
      \hline \hline
       & $\Delta m$ (MeV$/c^2$) \\
      \hline
      $B^{0}$ 		& $10.2  \pm 1.7  \ \pm 1.2   \pm 0.4$  \\
      $B^{+}$ 		& $10\ \ \, \pm 3  \ \ \ \ \, ^{+\,2}_{-\,1}\, \ \ \pm 2\ \ \,$ \\
      $B^{0}_s$ 	& $11.4  \pm 0.2  \ \pm 0.0   \pm 0.4$  \\
      \hline \hline
     \end{tabular} \\
\end{table}

\section{Summary}
Using the full CDF Run II data sample, we measure the masses and widths of $B^{**}_{(s)}$ mesons.
For the first time, we observe exclusively reconstructed $B^{**+}$ mesons and measure the width of the $B_{1}^{0}$ state.
The results are consistent with, and significantly more precise than previous determinations based on a subset of the present data~\cite{CDFmea,CDFsmea}, which are superseded. 
The results are also generally compatible with determinations by the D0~\cite{D0mea} and LHCb experiments~\cite{LHCbmea}.
The only exception is the remaining discrepancy with the D0 measurement of the mass difference between $B_1^{0}$ and $B_2^{0*}$ mesons, which increases to $4.2\sigma$.

The properties of the $B^{**0}$ and $B^{**+}$ states are within $2\sigma$ consistent with isospin symmetry.
The measured $B^{**0,+}$ masses are in agreement with the HQET predictions in Ref.~\cite{hqet4}.
The QCD string calculation in Ref.~\cite{qcdstring} matches data with a deviation of about 10~MeV/$c^2$.
The lattice calculation in Ref.~\cite{lattice1} predicts the $B_1$ mass accurately with a deviation of only 6~MeV/$c^2$, but is off by 35~MeV/$c^2$ for the $B_2^{*+}$ mass.
The heavy-quark symmetry and potential-model-based predictions in Ref.~\cite{hqs} and \cite{potmod1} are about 30~MeV/$c^2$ above and below the measured values, respectively.
Our measurement is consistent with the HQET predictions of the $B^{**}$ widths in Refs.~\cite{hqet2,hqet3} and the $\Gamma(B_2^*)$ prediction in Ref.~\cite{potmod2}.
The $B_s^{**0}$ masses are described by HQET calculations~\cite{hqet4,hqet6,hqs} within 3--6~MeV/$c^2$.
The lattice calculations in Ref.~\cite{lattice1} agree with the measurements within theoretical uncertainties.
The HQET prediction in Ref.~\cite{hqet2} and predictions based on chiral theory~\cite{chiral2}, potential models~\cite{potmod2}, and lattice calculations~\cite{lattice2} are about 30--60~MeV/$c^2$ too high.
The $B_s^{**0}$ width predictions by HQET~\cite{hqet2,hqs} are 1--2~MeV/$c^2$ above the measurements while the prediction of $\Gamma(B_{s2}^{*0})$ in Ref.~\cite{potmod2} agrees well with the experimental result.

We observe a previously-unseen charged and neutral $B\pi$ signal with a significance of $4.4 \sigma$. 
Interpreting it as a single state, referred to here as $B(5970)$, we measure the properties of the new resonance for charged and neutral $B\pi$ combinations and find them to be statistically consistent as expected by isospin symmetry.

\section{Acknowledgements}
We thank the Fermilab staff and the technical staffs of the participating institutions for their vital contributions. This work was supported by the U.S. Department of Energy and the National Science Foundation; the Italian Istituto Nazionale di Fisica Nucleare; the Ministry of Education, Culture, Sports, Science and Technology of Japan; the Natural Sciences and Engineering Research Council of Canada; the National Science Council of the Republic of China; the Swiss National Science Foundation; the A.P. Sloan Foundation; the Bundesministerium f\"ur Bildung und Forschung, Germany; the Korean World Class University Program, the National Research Foundation of Korea; the Science and Technology Facilities Council and the Royal Society, UK; the Russian Foundation for Basic Research; the Ministerio de Ciencia e Innovaci\'{o}n, and Programa Consolider-Ingenio 2010, Spain; the Slovak R\&D Agency; the Academy of Finland; the Australian Research Council (ARC); and the EU community Marie Curie Fellowship contract 302103.


\end{document}